\documentstyle[12pt]{article}
\def\beq{\begin{equation}}
\def\eeq{\end{equation}}

\def\beqn{\begin{eqnarray}}
\def\eeqn{\end{eqnarray}}
\relax
\begin{document}
\begin{titlepage}
\def\ba{\begin{array}{c}}
\def\ea{\end{array}}
\def\thefootnote{\fnsymbol{footnote}}
\vfill
\hskip 4.in  BNL-61012

\hskip 4.in  October, 1994
\vskip 1.2in
\begin{center}
{\large \bf INTRODUCTION TO  THE PHYSICS OF \\
\vskip .25in
HIGGS BOSONS}\\

\vspace{1 in}
{\bf S.~Dawson}\\
{\it  Physics Department,\\
               Brookhaven National Laboratory,\\  Upton, NY 11973}\\
\vspace{1 in}
\end{center}
\begin{abstract}
A basic introduction to the physics of the Standard Model Higgs boson is
given.  We discuss Higgs boson production in $e^+e^-$  and
hadronic collisions and survey search techniques at future
accelerators.  The Higgs bosons of the minimal SUSY model are
briefly considered.  Indirect limits from triviality arguments, vacuum
stability and precision measurements at LEP are also presented.

\end{abstract}

\vskip 1in {\it Lectures given at the 1994 Theoretical Advanced
Study Institute, Boulder, CO., May 30 -June 23, 1994}
\end{titlepage}

\clearpage

\tableofcontents
\clearpage
{\centerline {\bf List of Figures}}

Fig. 1~~Scalar potential with $\mu^2>0$.

Fig. 2~~Scalar potential with $\mu^2<0$.

Fig. 3~~Determination of $v$ from $\mu$ decay.

Fig. 4~~Feynman rules for the Standard Model Higgs boson.

Fig. 5~~Higgs production through $e^+e^- \rightarrow Z H$.

Fig. 6~~Branching ratio for $Z\rightarrow H e^+e^-$.

Fig. 7~~Total cross section for $e^+e^- \rightarrow ZH$.

Fig. 8~~Higgs boson production through gluon fusion with a quark loop.

Fig. 9~~Form factor for gluon fusion, $\mid I\biggl(
{ 4M_T^2\over M_H^2}\biggr)\mid^2$.

Fig. 10~~Gluon-gluon luminosity in $pp$ (or $ p
{\overline p}$) collisions at $\sqrt{s}=1.8~TeV$ and $14~TeV$.

Fig. 11~~Cross section for Higgs production
from gluon fusion
at the LHC from a top quark with $M_T=170~GeV$.

Fig. 12~~Effective $ggH$ vertex.

Fig. 13~~Radiatively corrected cross section, $\sigma_{\rm TOT}$,
 for gluon
fusion production of the Higgs boson at the LHC.

Fig. 14~~Branching ratios for the intermediate mass Higgs boson.

Fig. 15~~Irreducible backgrounds to $H\rightarrow \gamma\gamma$;
$q {\overline q} \rightarrow \gamma\gamma$ and $g g \rightarrow
\gamma\gamma$.

Fig. 16~~Expected $M_{\gamma \gamma}$ signal for $H\rightarrow \gamma
\gamma$ signal above the irreducible $\gamma \gamma$ background
for $M_H=110~GeV$ at the LHC using the ATLAS detector.  This figure
is from Ref. [18].

Fig. 17~~Production of the Higgs boson through $q \overline{q}^{\prime}
\rightarrow W H$.

Fig. 18~~Higgs boson production at the Tevatron.

Fig. 19~~Higgs boson production from $WW$ fusion.

Fig. 20~~$H\rightarrow W^+_LW^-_L$.

Fig. 21~~Amplitude for a polarized $W$ to scatter
from a quark into the final state $X$.

Fig. 22~~Quark-quark scattering by vector boson exchange.

Fig. 23~~Luminosity of $W$ boson pairs in the proton at the LHC.
  This figure is from Ref. [32].

Fig. 24~~QCD corrections to the effective $W$ approximation.

Fig. 25~~Longitudinal and transverse gauge boson contributions (solid) to
$pp\rightarrow ZZ$ through intermediate $W^+W^-$ and $ZZ$
interactions for $M_H=500~GeV$ at the SSC.  This figure is
from Ref. [36].

Fig. 26~~Feynman diagrams contributing to $ZZ$ production
from $WW$ fusion.

Fig. 27~~Various contributions to $W^+W^-$ production.  The dot-
dashed line is the background from $q \overline{q}\rightarrow
W^+W^-$, the dashed is vector boson scattering and the
dotted line is the contribution from $s$-channel Higgs exchange only.
The solid line is the total rate.
This figure is from Ref. [36].

Fig. 28~~Comparison of the Effective $W$ approximation (dashed) with
the exact numerical calculation(solid) for $ZZ$ production
from $W^+W^-$ fusion with $\mid \eta_Z \mid < 1.5$.
This figure is from Ref. [37].

Fig. 29~~Diagrams not contained in the effective $W$ approximation.

Fig. 30~~Processes contributing to Higgs production at the LHC.

Fig. 31~~$pp\rightarrow ZZ\rightarrow 4$
leptons for $M_H=800~GeV$
at the LHC with the ATLAS detector applying a series of cuts described
in the text.  This figure is from Ref. [18].

Fig. 32~~$pp\rightarrow
ZZ\rightarrow l^+l^-\nu{\overline \nu}$ for $M_H=500~GeV$ with
$\int{\cal L}= 10^{4} pb^{-1}$ at the LHC using the ATLAS
detector.  This figure is from Ref. [18].

Fig. 33~~Feynman rules for Higgs-Goldstone boson interactions.

Fig. 34~~Goldstone boson scattering, $\omega^+\omega^-\rightarrow
\omega^+\omega^-$.

Fig. 35~~Effects of anomalous gauge boson interactions in
$e^+e^-$ interactions.

Fig. 36~~Strong interaction effects in hadronic collisions.
The dotted line is the number of $W^+_LW^+_L$ events at the LHC from
vector boson fusion with $\int {\cal L}=10^{40}/cm^2$ for
$.5 < M_{WW}< 1~TeV$.
We have assumed $L_2=-L_1$ with all other $L_i=0$  and a
renormalization scale $\mu=1.5~TeV$.

Fig. 37~~Argand diagram showing unitarity condition on scattering
amplitudes.

Fig. 38~~Upper bound on the Higgs mass in a pure $\Phi^4$ theory.
The solid line is the result of Eq. 115.
The dashed line is $\Lambda=v$ and the dotted line
$\Lambda=M_H^{\rm max}$.

Fig. 39~~Lattice gauge theory bound on the Higgs boson mass.  This figure
is from Ref. [65].

Fig. 40~~Mass of the lightest Higgs boson of a SUSY model
(in GeV)  in terms
of $\tan\beta$ and $M_A$ for $M_T=150~GeV$.  This figure is from Ref. [80].

Fig. 41~~Couplings of the lightest SUSY Higgs boson, $H_1$ to
the $Z$.

Fig. 42~~Limits on SUSY parameter space from ALEPH [80].
\clearpage
The search for the Higgs boson has become the holy grail of
all particle accelerators. In the simplest version of the
electroweak theory, the Higgs boson serves both to give
the $W$ and $Z$ bosons their masses and to give the fermions
mass.   It is thus a vital part of the theory.
   In these lectures, we will introduce
the Higgs boson of the Standard Model of electroweak interactions.\cite{hhg}
We discuss the production of the Higgs boson in both $e^+e^-$ collisions and
hadronic interactions and survey search techniques
in the various Higgs mass ranges.\cite{bag}

Section 1 contains a derivation of the Higgs mechanism, with particular
emphasis on the choice of gauge.  In Section 2 we
 discuss  Higgs production in $e^+e^-$ collisions and
describe the current LEP bounds on the Higgs mass.  Hadronic
production of the Higgs boson through gluon fusion
and potential discovery channels at the LHC are the subjects
of Section 3.  Section 4 contains a derivation of the effective
$W$ approximation and a discussion of Higgs production through vector
boson fusion at the LHC.

Suppose the Higgs boson is not discovered in an
$e^+ e^-$ collider or at the LHC?  Does this mean the
Standard Model with a Higgs boson must be abandoned?  In
Section 5, we discuss the implications of a very heavy Higgs boson,
($M_H>> 700~GeV$).  Finally, in Section 6, we present indirect
limits on the Higgs boson mass from triviality arguments,
 vacuum stability
and precision $e^+e^-$ measurements.  Section 7 contains
a list of some of the objections which many theorists
 have to the minimal standard model
with a single Higgs boson.  One of the most popular alternatives to the
minimal Standard Model is to make the theory supersymmetric, which is
discussed in Section 8.  We present a lightning review of those
aspects of supersymmetric theories which are
relevant for Higgs boson phenomenology.
We end with some conclusions in
Section 9.

\section{The Higgs Mechanism}
\subsection{Abelian Higgs Model}

The central question of electroweak physics is :``Why are the $W$ and $Z$
boson masses non-zero?".  The measured values,
$M_W=80~GeV$ and $M_Z=91~GeV$ are
 far from zero and cannot be considered as
small effects.
  To see that this is a problem, we consider
a $U(1)$ gauge theory with a single gauge field, the photon.  The Lagrangian
is simply\cite{quigg}
\beq
{\cal L}=-{1\over 4} F_{\mu \nu}F^{\mu\nu},
\eeq
where
\beq F_{\mu\nu}=\partial_\nu A_\mu-\partial _\mu A_\nu.
\eeq
The statement of local $U(1)$ gauge invariance is that the Lagrangian
is invariant under the transformation:$A_\mu(x)
\rightarrow A_\mu(x)-
\partial_\mu \eta(x)$ for any $\eta$.   Suppose we now add a
mass term to the Lagrangian,
\beq
{\cal L}=-{1\over 4} F_{\mu \nu}F^{\mu\nu}+{1\over 2}m^2 A_\mu A^\mu
{}.
\eeq
It is easy to see that the mass term violates the local gauge invariance.
Hence it is gauge invariance that requires the photon to be massless.

We can extend the model by adding a single complex scalar field,
\beq
\phi\equiv {1\over \sqrt{2}}(\phi_1 + i \phi_2).
\eeq
The Lagrangian is now,
\beq
{\cal L}=-{1\over 4} F_{\mu\nu} F^{\mu\nu}+\mid D_\mu\phi\mid^2
-V(\phi),
\eeq
where
\beqn D_\mu & =&\partial_\mu -i e A_\mu \nonumber \\
V(\phi) &=& \mu^2 \mid \phi\mid^2+\lambda(\mid \phi\mid^2)^2 . \eeqn
$V(\phi)$ is the most general renormalizable potential allowed by
the $U(1)$ invariance.

This Lagrangian is invariant under $U(1)$ rotations, $\phi
\rightarrow e^{i\theta}\phi$ and under local gauge transformations:
\beqn
A_\mu(x) &\rightarrow & A_\mu(x)-\partial _\mu \eta(x) \nonumber \\
\phi(x) &\rightarrow & e^{-i e \eta(x)} \phi(x).\\
\eeqn

There are two possibilities for the theory.\footnote{We
assume $\lambda>0$.}  If $\mu^2>0$ then the
potential has the shape shown in Fig. 1 and preserves the symmetry
of the Lagrangian.
The state of lowest energy is that with $\phi=0$, the vacuum state.
The theory is simply quantum electrodynamics
with a massless photon and a charged
scalar field $\phi$ with mass $\mu$.

The alternative scenario
 is more interesting.  In this case $\mu^2<0$ and the
potential can be written as
\beq V(\phi)=-\mid \mu\mid^2 \mid \phi\mid^2 +
\lambda (\mid \phi\mid^2)^2,
\eeq
which has the Mexican hat shape shown in Fig. 2.  In  this case
the minimum energy state is not at $\phi=0$ but rather at
\beq
\langle \phi\rangle=\sqrt{-{\mu^2\over 2 \lambda}}.
\eeq
$\langle \phi\rangle$ is called the vacuum expectation value (VEV)
of $\phi$.
Note that the direction in which the vacuum is chosen is
arbitrary, but it is conventional to choose it to lie along
the direction of the real part of $\phi$ as shown in Fig. 2.
 It is convenient to write $\phi$ as
\beq \phi\equiv {1\over \sqrt{2}}e^{i {\chi\over v}}
\biggl(v+H\biggr),
\label{phidef}
\eeq
where $\chi$ and $H$ are real fields which have no VEVs.
If we substitute Eq.~\ref{phidef}  back into the original Lagrangian we find
the interactions in terms of the fields with no VEVs,
\beqn
{\cal L} & = &
-{1\over 4} F_{\mu\nu}F^{\mu\nu} - e v A_{\mu}\partial^{\mu}\chi
+{e^2 v^2\over 2} A_{\mu} A^\mu \nonumber \cr
&&+{1\over 2} \biggl( \partial_\mu H \partial^\mu H
+ 2 \mu^2 H^2\biggr)
+{1\over 2} \partial_\mu\chi\partial^\mu\chi\nonumber \\
&&+(H,~\chi {\rm ~interactions}).
\label{ints}
\eeqn
Eq.~\ref{ints}
 describes a theory with a photon of mass $M_A=ev$, a scalar field
$H$ with mass-squared $-2 \mu^2>0$, and a massless scalar field $\chi$.
The mixed $\chi-A$ propagator is confusing however.  This term can be
removed by making a gauge transformation:
\beq
A^\prime_\mu\equiv A_\mu -{1\over e v} \partial_\mu\chi.
\label{gauget}
\eeq
After making the gauge transformation of Eq.~\ref{gauget}
 the $\chi$ field disappears from the theory and we
say that it has been ``eaten'' to give the photon mass.
The $\chi$ field is often called a Goldstone
boson.\cite{gold}
 In the gauge of Eq. \ref{gauget}
the particle content of the theory
 is  apparent; a massive photon and
a scalar field $H$.
 The field $H$ is called a Higgs boson.
The Higgs mechanism can be summarized by saying that the spontaneous
breaking of a gauge theory by a non-zero VEV results in the
disappearance of a Goldstone boson and its
transformation into the longitudinal component of a massive
gauge boson.

It is instructive to count the degrees of freedom (dof).  Before the
spontaneous symmetry breaking there was a massless photon (2 dof) and
a complex scalar field (2 dof) for a total of
 4 degrees of freedom.\footnote{Massless gauge fields have 2 transverse
degrees of freedom, while a massive gauge field has an additional
longitudinal degree of freedom.}
After the spontaneous symmetry breaking there is a massive photon
(3 dof) and a real scalar, $H$, (1 dof) for the same total degrees
of freedom.

At this point let us make an aside about the gauge dependance of these
results.  The gauge choice above with the transformation
$A_\mu^\prime=A_\mu -{1\over e v }\partial_\mu\chi$ is called the
unitary gauge.  This gauge has the advantage that the particle
spectrum is obvious and there is no $\chi$ field.
The unitary gauge however has the disadvantage that the vector
propagator, $\Delta_{\mu\nu}(k)$,
 has bad high energy behaviour,
\beq
\Delta_{\mu\nu}(k)=-{i\over k^2-M_A^2}\biggl( g_{\mu\nu}-{k^{\mu}k^{\nu}
\over M_A^2}\biggr).
\eeq
In the unitary gauge, scattering cross sections have contributions which
grow with powers of $k^2$ (such as $k^4$, $k^6$, etc.) which cannot
be removed by the conventional mass, coupling constant, and wavefunction
renormalizations.
More convenient gauges are the so-called $R_{\xi}$ gauges
which are obtained by adding the gauge fixing term to
the Lagrangian,\cite{ablee}
\beq {\cal L}_{GF}=-{1\over 2 \xi}\biggl(\partial_\mu A^\mu+\xi
e v \chi\biggl)^2.
\label{rtsi}
\eeq
Different choices for $\xi$ correspond to different gauges.
In the limit $\xi\rightarrow \infty$ the unitary gauge is recovered.
Note that the cross term in Eq. ~\ref{rtsi} cancels the mixed
$\chi \partial_\mu A^\mu$ term of Eq. \ref{ints}.
The gauge boson propagator in $R_\xi$ gauge is given by
\beq \Delta_{\mu\nu}(k)=-{i\over k^2-M_A^2}
\biggl( g_{\mu\nu}-{(1-\xi)k_{\mu}k_{\nu}\over k^2-\xi M_A^2}\biggr).
\eeq
 In the $R_\xi$ gauges the $\chi$ field is part of the spectrum
and has mass $M_{\chi}^2=\xi M_A^2$.  Feynman gauge corresponds to the
choice $\xi=1$  and has massive Goldstone bosons,
while Landau gauge has $\xi=0$ and  massless Goldstone bosons with no
coupling to the physical Higgs boson.
The Landau gauge is often the most convenient for calculations involving
the Higgs boson since there is no coupling to the unphysical $\chi$ field.

\subsection{Weinberg-Salam Model}

It is now straightforward to obtain the usual Weinberg-Salam model
of electroweak interactions.\cite{ws}
   Here we present a quick overview
of the model with emphasis on those aspects relevant for Higgs physics.
Further details can be found in the lectures of Peccei at this
school.\cite{pec}
The Weinberg- Salam model is an $SU(2)_L \times U(1)_Y$ gauge theory containing
$3$ $SU(2)$ gauge bosons, $W_\mu^i$, and one $U(1)$
gauge boson, $B_\mu$, with a kinetic energy term,
\beq
{\cal L}_{\rm KE} =-{1\over 4}W_{\mu\nu}^i W^{\mu\nu i}
-{1\over 4} B_{\mu\nu} B^{\mu\nu}
\eeq
where
\beqn
W_{\mu\nu}^i&=& \partial_\nu W_\mu^i-\partial _\mu W_\nu^i
+g \epsilon^{ijk}W_\mu^j W_\nu^k
\nonumber \\
B_{\mu\nu}&=&\partial_\nu B_\mu-\partial_\mu B_\nu\quad .
\eeqn
Coupled to the gauge fields is a complex scalar
doublet
\beq
\Phi={1\over \sqrt{2}}
\left(\begin{array}{c}
\phi_1+i\phi_2  \\
H+i\phi_0   \end{array}\right) \quad ,
\eeq
with a  scalar potential is given by
\beq
 V(\Phi)=\mu^2 \mid \Phi^\dagger\Phi\mid +\lambda
\biggl(\mid \Phi^\dagger \Phi\mid\biggr)^2\quad ,
\label{wspot}
\eeq
($\lambda>0$).
This is the most general renormalizable and $SU(2)$ invariant
potential allowed.

Just as in the Abelian model of Section 1.1, the state of minimum
energy for $\mu^2<0$ is not at $0$ and the scalar field develops
a VEV.
The direction of the minimum in $SU(2)_L$ space is not determined
since the potential depends only on $\Phi^\dagger \Phi={1\over 2}(
\phi_1^2+\phi_2^2+H^2+\phi_0^2)$
and we choose
\beq
\langle \Phi\rangle
= {1\over\sqrt{2}} \left(\begin{array}{c}
 0   \\
 v   \end{array}\right)\quad .
\label{vevdef}
\eeq
With this choice the scalar doublet has $U(1)_Y$ charge
(hypercharge) $Y_\Phi=1$ and the electromagnetic charge is
$Q=T_3 +{Y\over 2}$.  Therefore,
\beq
Q \langle \Phi\rangle
= 0
\eeq
and electromagnetism is unbroken by the scalar VEV.
The VEV of Eq. \ref{vevdef} hence yields the desired symmetry breaking
scheme, $SU(2)_L\times U(1)_Y\rightarrow U(1)_{EM}$.

It is now straightforward to see how the Higgs mechanism
generates mass for the $W$ and $Z$ gauge bosons.  The
contribution of the scalar fields to the Lagrangian is,
\beq
{\cal L}_s=(D^\mu \Phi)^\dagger (D_\mu \Phi)-V(\Phi)
\eeq
where\footnote{The $\tau_i$ are the Pauli matrices with $Tr(\tau_i\tau_j)
=2\delta_{ij}$. Different choices for the gauge kinetic energy
and the covariant derivative depend on whether $g$ and $g^\prime$
are chosen positive or negative.  There is no physical consequence
of this choice.}
\beq
D_\mu=\partial_\mu +i {g\over 2}\tau\cdot W_\mu+i{g^\prime\over 2}
B_\mu.
\eeq
In unitary gauge the scalar field can be written as
\beq
\Phi={1\over \sqrt{2}}\left(\begin{array}{c}  0 \\
 v+H\end{array}\right)
\eeq
which gives the contribution to the gauge boson masses
from the scalar kinetic energy term,
\beq
{1\over 2} (0 \quad v )
\biggl({1\over 2}g \tau\cdot W_\mu
+{1\over 2} g^\prime B_\mu
\biggr)^2 \left(\begin{array}{c}  0 \\  v \end{array}
\right).
\eeq
Hence the gauge fields obtain a mass from the Higgs mechanism:
\beqn
W^{\pm}_\mu&=&
{1\over \sqrt{2}}(W_\mu^1 \mp i W_\mu^2)\nonumber \\
Z^\mu&=& {-g^\prime B_\mu+ g W_\mu^3\over \sqrt{g^2+g^{\prime~2}}}
\nonumber \\
A^\mu&=& {g B_\mu+ g^{\prime} W_\mu^3\over \sqrt{g^2+g^{\prime~2}}}.
\eeqn
The coupling constants satisfy the usual relations,
\beqn
e&=& g \sin\theta_W \nonumber \\
e&=& g^\prime \cos\theta_W
\eeqn
and the masses are given by
\beqn
M_W^2 &=& {1\over 4} g^2 v^2\nonumber \\
M_Z^2 &=& {1\over 4} (g^2 + g^{\prime~2})v^2\nonumber \\
M_\gamma& = & 0.
\eeqn
It is again
instructive to count the degrees of freedom after the spontaneous
symmetry breaking has occurred.  We began with a complex scalar field
$\Phi$ with four degrees of freedom, a massless $SU(2)$ gauge field,
$W_i$, with six degrees of freedom and a massless $U(1)$ gauge field,
$B$, with 2 degrees of freedom for a total of $12$.  After the
spontaneous symmetry breaking there remains a physical real scalar field
$H$ ($1$ degree of freedom),  massive $W$ and $Z$ fields ($9$
degrees of freedom), and a massless photon ($2$ degrees of freedom).
We say that the scalar degrees of freedom have been ``eaten'' to
give the $W$ and $Z$ gauge bosons their longitudinal components.

Just as in the case of the Abelian Higgs model, if we go to
a gauge other than unitary gauge, there will be Goldstone
bosons in the spectrum and the scalar field can be written,
\beq
\Phi={ e^{i{\omega\cdot\tau\over v}}\over \sqrt{2}}
\left(\begin{array}{c}  0 \\
 v+H\end{array}\right).
\eeq
In the Standard Model, there are three Goldstone bosons,
${\vec \omega}=(\omega^\pm,z)$, with masses $M_W$ and $M_Z$ in
the Feynman gauge.  These Goldstone bosons will play an important
role in our understanding of the physics of a very heavy Higgs
boson, $M_H>1~TeV$, as we will discuss in Section 5.1.

In addition to giving the $W$ and $Z$ bosons their masses, the Higgs
boson is also used to give the fermions mass. The
gauge invariant  Yukawa coupling of the
Higgs boson to fermions is
\beq
{\cal L}_f=-\lambda_d {\overline Q}_L \Phi d_R + h.c.\quad ,
\eeq
where the left handed $SU(2)$ fermion doublet  is
\beq
Q_L=
\left(\begin{array}{c}
u\\ d \end{array}\right)_L
{}.
\eeq
This gives the effective coupling
\beq
\lambda_d {1\over\sqrt{2}}
({\overline u}_L,~ {\overline d}_L)\left(
\begin{array}{c}  0 \\
v+ H \end{array} \right) d_R + h.c.
\eeq
which can be seen to yield a mass term for the down quark if
we make the identification
\beq
\lambda_d = {m_d \sqrt{2}\over v}.
\eeq
In order to generate a mass term for the up quark we use the
fact that $\Phi^c \equiv - i \tau_2 \Phi^*$ is an $SU(2)$
doublet and  we can write the $SU(2)$ invariant coupling
\beq
\lambda_u {\overline Q}_L \Phi^c u_R + h.c.
\eeq
which generates a mass term for the up quark.  Similar couplings
can be used to generate mass terms for the charged leptons.
For the multi-family case, the Yukawa couplings, $
\lambda_d$ and $\lambda_u$, become $N_F \times N_F$ matrices
(where $N_F$ is the number of families).  Since the mass matrices
and Yukawa matrices  are proportional, the interactions of the
Higgs boson with the fermion mass eigenstates are flavor diagonal.
That is, the Higgs boson does not mediate flavor changing interactions.

By expressing the fermion kinetic energy in terms of the gauge
boson mass eigenstates, the charged and neutral weak current interactions
can be found.
The parameter $v$ can be determined from
the charged current for $\mu$ decay,
$\mu\rightarrow e {\overline \nu}_e \nu_\mu$, as shown in
Fig. 3.
Since the momentum carried by the $W$ boson is of order $m_\mu$ it
can be neglected in comparison with $M_W$ and we make the identification
\beq
{G_F\over \sqrt{2}}={g^2\over 8 M_W^2}={1\over 2 v^2},
\eeq
which gives the result $v^2=(\sqrt{2} G_F)^{-1} = (246~GeV)^2$.

One of the most important points about the Higgs mechanism is
that all of the couplings of the Higgs boson to fermions and
gauge bosons are completely determined in terms of coupling
constants and fermion masses. The potential of Eq. \ref{wspot}
had two free parameters, $\mu$ and $\lambda$.  We can trade
these for
\beqn
v^2&=&-{\mu^2\over 2 \lambda}\nonumber \\
M_H^2&=& 2 v^2 \lambda.
\eeqn
  There are no free adjustable
parameters and so Higgs production  and decay processes
can be computed unambiguously in terms of the Higgs mass.  In Fig. 4
we give a complete set  of Feynman rules for the couplings of the
Higgs boson.  Note that for $M_H>>v$, the self couplings of the Higgs
boson become strong.

\section{Higgs Production in $e^+ e^-$ Colliders}

Since the Higgs boson coupling to the electron is very
small, $\sim m_e/v$, its dominant production mechanism in
$e^+e^-$ collisions is the so called ``Bjorken Mechanism''
shown in Fig. 5.\cite{bj}
An estimate of the size of Higgs production can be found from
the decay $Z\rightarrow H f {\overline f}$ for a massless Higgs boson:
\beqn
{BR(Z\rightarrow H f \overline{f})\over
BR(Z\rightarrow f \overline{f})}\mid_{M_H=0}
&=&{g^2\over 192\pi^2 \cos \theta_W^2}
\biggl[\biggl(6-{\Gamma_Z^2\over 2 M_Z^2}\biggr)\log
\biggl({\Gamma_Z^2+M_Z^2\over \Gamma_Z^2}\biggr)\nonumber \\
&& \qquad
+{12\Gamma_Z\over M_Z}\tan^{-1}\biggl({M_Z\over \Gamma_Z}\biggr)
-{23\over 2}\biggr]\nonumber \\
&\sim& 10^{-2},
\label{zhff}
\eeqn
where $\Gamma_Z$ is the total $Z$ boson decay width.
We can see from Eq. \ref{zhff}
that Higgs boson production in $Z$ decays
can never be more than a few percent effect.

The Higgs boson has been searched for in $e^+e^-$ collisions at
the LEP collider, which has $\sqrt{s}=M_Z$.
The Higgs is produced through the mechanism of Fig. 5, but with
the final $Z$ off-shell  and decaying to a lepton pair.
    The primary decay
mechanism used is $Z^*\rightarrow H l^+ l^-$ whose
branching ratio is shown in Fig. 6.
  The decay $Z^*\rightarrow H \nu {\overline \nu}$
is also useful since the branching ratio is 6 times larger
than that of $Z\rightarrow H e^+ e^-$.
  The strategy is
to search for each range of Higgs mass separately by looking for the
relevant Higgs decays.
For example, a light
Higgs boson, $M_H< 2 m_e$ , necessarily decays to $2$
photons.  For $M_H\sim 1~ MeV$, the Higgs
lifetime gives $c \tau\sim 10^3~cm$
and so the Higgs boson is long lived and escapes the detector
without interacting.  In this case the relevant reaction is
$e^+e^-\rightarrow Z\rightarrow l^+ l^- H$ and  the signal
is $l^+ l^-$ plus missing energy from the undetected Higgs boson.
For each mass region, the appropriate Higgs decay channels are searched for.
When the Higgs becomes heavier than twice the $b$ quark mass it
decays primarily to $b \overline{b}$ pairs and the signal
is then  $e^+e^-\rightarrow Z\rightarrow l^+ l^- H\rightarrow
l^+l^-+{\rm jets}$.  By a systematic study of all Higgs boson
masses and decay channels,
 the LEP experiments have found the limit\cite{pdg}\footnote{
Note that there is no region where light Higgs boson masses are allowed.
The LEP limits thus obviate early studies of mechanisms such as $K\rightarrow
\pi H$ or $B\rightarrow \pi H$.}
\beq
M_H> 58~GeV.
\eeq
Since the LEP experiments have on the order of one million $Z$'s
we see from Fig.~6 that the number of Higgs events expected for
higher masses is quite small and this limit is not expected to
be significantly improved by future running.\cite{moro}

In the future, LEPII will run at an energy somewhere above
$\sqrt{s}\sim 175~GeV$ and so will look for the process
$e^+e^-\rightarrow Z H$ shown in Fig. 5.  The only difference between this
process and the
 searches at LEP is that now the final state $Z$ can be on-shell.
The cross section for this process at a center- of- mass energy $s$
is,\cite{sigee}
\beq
\sigma(e^+e^-\rightarrow H Z)={\pi\alpha^2\sqrt{\lambda}
(\lambda+12 s M_Z^2)[1+(1-4 \sin^2\theta_W)^2]\over
192 s^2 \sin^4\theta_W \cos^4 \theta_W (s-M_Z^2)^2}
\eeq
where $\lambda\equiv(s-M_H^2-M_Z^2)^2-4 M_H^2 M_Z^2$. (In the
center of mass, the momentum of the outgoing $Z$ is
${\sqrt{\lambda}\over
2\sqrt{s}}$).  From this we can see that the cross section
peaks at an energy $\sqrt{s}\sim M_Z+\sqrt{2}M_H$.  This is
a very clean production channel with little background and
LEPII is expected to be able to explore the Higgs mass region
up to the kinematic limit, $M_H < \sqrt{s}-M_Z\sim {\cal O}(80~GeV)$.

In Fig. 7, we show the total cross section for $e^+e^-\rightarrow
Z H$ as a function of $\sqrt{s}$ for fixed $M_H$.  If we
demand  $40$ $ZH$ events with $Z\rightarrow e^+e^- +
\mu^+\mu^-$ in $1000~pb^{-1}$ to discover the
Higgs in this channel, then we require $\sigma>.7~pb$ which implies
that LEPII will be sensitive to $M_H < 80~GeV$,
which is consistent with our estimate that LEPII will
reach the kinematic limit.

\section{Higgs Production in Hadronic Collisions}
\subsection{Gluon Fusion}
We turn now to the production of the Higgs boson in $pp$ or $p \overline{p}$
collisions.  Since the coupling of a Higgs boson to an up quark
or a down quark is proportional to the quark mass, this coupling
is very small.  The primary production mechanism for a Higgs boson
in hadronic collisions is through gluon fusion, $g g \rightarrow H$,
which is shown in Fig. 8.  The loop contains all quarks with mass $m$.
(In extensions of the standard model, all massive colored particles
run in the loop.)
To evaluate the diagram of Fig. 8, we use dimensional regularization
in $n=4-2\epsilon$ dimensions.
For a fermion of mass $m$ in the loop the amplitude
given by the diagram of Fig. 8.
is
\beq
i {\cal A} = - (-i g_s)^2 Tr (T_A T_B) \biggl({- i m\over v}
\biggr)
\int {d^n k\over (2 \pi)^n} {T^{\mu\nu}\over D} (i)^3
\epsilon_\mu(p)\epsilon_\nu(q)
\label{ggh}
\eeq
where the overall minus sign is due to the closed fermion loop.\footnote{
$\epsilon_\mu(p)$ are the transverse gluon polarizations.}
The denominator is $D= (k^2-m^2)[(k+p)^2-m^2][(k-q)^2-m^2]$. The
usual method of Feynman parameterization can be used to combine
the denominators,
\beq
{1\over A B C}= 2 \int_0^1dx \int^{1-x}_0{d y
\over [A x +By +C(1-x-y)]^3}
\eeq
and so the denominator becomes,
\beq
{1\over D}\rightarrow
2 \int dx ~dy {1\over [k^2-m^2 + 2 k \cdot (px-qy)]^3}.
\eeq
Shifting the integration momenta, $k^{\prime}= k + p x - q y$,
the denominator takes the form
\beq
{1\over D}\rightarrow 2 \int dx~dy {1\over [ k^{\prime~2}
- m^2 +M_H^2 x y]^3}.
\eeq
The numerator of Eq. ~\ref{ggh} is also easily evaluated
\beqn
T^{\mu\nu}&=& Tr\biggl[
(k+m)\gamma^\mu
(k+p+m)(k-q+m)\gamma^\nu)\biggr]
\nonumber \\
&=& 4 m\biggl[ g^{\mu\nu}(m^2-k^2-{M_H^2\over 2})
+4 k^{\mu}k^{\nu} + p^\nu q^\mu\biggr]
\eeqn
where we have used the fact that for transverse gluons,
$\epsilon(p)\cdot p=0$ and so terms proportional
to $p_\mu$ or $q_\nu$ can be dropped.
We now shift momenta, drop terms linear in $k^\prime$ from the
numerator and use the relation
\beq
\int d^n k^{\prime} {k^{\prime\mu}k^{\prime\nu}\over (k^{\prime 2}
-C)^m}={1\over n} g^{\mu\nu}
\int d^nk^{\prime} {k^{\prime 2}\over (k^{\prime 2}-C)^m}
\eeq
to write the amplitude in the form
\beqn
i {\cal A}&=&
-{2 g_s^2 m^2\over v}
\delta_{AB} \int{d^n k^{\prime}
\over (2 \pi)^n}\int dx dy \biggl\{ g^{\mu\nu} \biggl[m^2 +k^{\prime 2}
\biggl({4\over n}-1\biggr)+M_H^2(xy-{1\over 2})\biggr]
\nonumber \\
&& +p^\nu q^\mu(1-4 x y)\biggr\}{2 dx dy\over
(k^{\prime 2}-m^2+M_H^2 x y)^3}
\epsilon_\mu(p)\epsilon_\nu(q).
\label{aggans}
\eeqn

The integral of Eq. \ref{aggans} can easily be done using the well
known formulas of dimensional regularization\cite{dim}
\beqn
\int {d^n k^{\prime}\over (2 \pi)^n}
{k^{\prime 2}\over (k^{\prime 2}-C)^3}&=&
{i\over 32 \pi^2}(4 \pi)^\epsilon
{\Gamma(1+\epsilon)\over \epsilon}(2-\epsilon)C^{-\epsilon}
\nonumber \\
\int {d^n k\over (2 \pi)^n}
{1\over (k^{\prime 2}-C)^3}&=&-
{i\over 32 \pi^2}(4 \pi)^\epsilon
\Gamma(1+\epsilon)C^{-1-\epsilon} .
\eeqn
We find the well known result\cite{glue}
\beq
{\cal A}(g g \rightarrow H)
=- {\alpha_s m^2 \over  \pi v}
\delta_{AB}\biggl(g^{\mu\nu}{M_H^2\over 2}-p^\nu q^\mu
\biggr)\int dx dy \biggl({1-4 x y
\over m^2-M_H^2 x y}\biggr)
\epsilon_\mu(p)\epsilon_\nu(q).
\label{gluans}
\eeq
(Note that we have multiplied by $2$ in Eq. \ref{gluans} to include
the diagram where the gluon legs are crossed.)
The Feynman integral of Eq.~\ref{gluans} can easily be performed to
find  an analytic result if desired.

It is particularly interesting to consider the case when
the fermion in the loop is much more massive than the
Higgs boson, $m>>M_H$.  In this case we find,
\beq
{\cal A}(gg\rightarrow H)\longrightarrow_{m>>M_H}
-{\alpha_s\over 3 \pi v}\delta_{AB}
\biggl(g^{\mu\nu}{M_H^2\over 2}- p^\nu q^\mu
\biggr)\epsilon_\mu(p)\epsilon_\nu(q).
\label{heavym}
\eeq
We see that the production process $g g \rightarrow H$ is
independent of the mass of the heavy fermion in the loop
in the limit $m>>M_H$.
Hence it counts the number of heavy generations and is a window into
 new physics at scales much above the energy being probed.  This is
a contradiction of our intuition that heavy particles should
decouple and not affect the physics at lower energy.  The
reason the heavy fermions do not decouple is, of course, because
the Higgs boson couples to the fermion mass.\cite{apc}

Resonant production of a heavy Higgs can be found from the
standard formula:\cite{cahn}
\beq
{\hat\sigma}(g g\rightarrow H)={\pi^2\over 8  M_H^3}
\Gamma(H\rightarrow g g)\delta(1-{M_H^2\over {\hat s}})  .
\eeq
It is straightforward to obtain our parton level result:
\beq {\hat \sigma}(g g \rightarrow H)=
{\alpha_s^2\over 64 \pi v^2}M_H^2 \mid I\biggl({M_H^2\over m^2}
\biggr)\mid^2 \delta({\hat s}-M_H^2)
\eeq
where $\sqrt{{\hat s}}$ is the energy in the gluon -gluon
center of mass and the integral $I$ is defined by
\beq
 I(a) \equiv \int_0^1 dx \int^{1-x}_0  dy{1-4xy\over 1-a x y}.
\label{intdef}
\eeq
In Fig. 9 we plot $I(a)$ and see that it goes quickly to its
large $a$ value.
Numerically, the heavy fermion mass limit is an extremely good
approximation even for $m\sim M_H$.
{}From this plot we can also see that the contribution
of light quarks to gluon fusion of the Higgs boson
 is irrelevant.  In fact we
have,
\beq I(a)\longrightarrow_{a\rightarrow\infty}
\sim -{1\over 2a} \log^2(a) . \eeq
Therefore, for the Standard Model, only the top quark is
numerically important when computing Higgs boson production from gluon
fusion.

To find the physical cross section we must integrate with the distribution
of gluons in a proton,\cite{scotttasi}
\beq
\sigma(pp\rightarrow H)=\int dx_1 dx_2 g(x_1) g(x_2)
{\hat \sigma}(gg\rightarrow H),
\eeq
where $g(x)$ is the distribution of gluons in the proton.
Better numerical convergence is obtained if we make the transformation
of variables, $x_1\equiv \sqrt{\tau} e^y$,
$x_2\equiv \sqrt{\tau} e^{-y}$, and  $\tau\equiv M_H^2/{\hat s}$.
The result is then,
\beq
\sigma(pp\rightarrow H) = {\alpha_s^2\over 64 \pi v^2}M_H^2
\mid I\biggl({M_H^2\over m^2}\biggr)\mid^2 {1\over s}{d {\cal L}
\over d \tau}
\eeq
where the gluon- gluon luminosity is defined
\beq
 {\tau\over s}{d {\cal L}
\over d \tau}\equiv {1\over s}\int^{-\log(\sqrt{\tau})}_{\log(\sqrt{\tau})}
 dy~ g\biggl(\sqrt{\tau}e^y
\biggr)g\biggl(\sqrt{\tau}e^{-y}\biggr).
\eeq
The gluon-gluon luminosity is shown in Fig. 10 for Tevatron
and LHC energies.
We see that the luminosity increases rapidly with energy.

It is straightforward to use the results given above to find the
cross section for Higgs production at the LHC (Large Hadron
Collider), a planned $pp$ collider at CERN  with an energy of $14~TeV$.
We show in Fig. 11 the cross section for producing a Higgs boson at
the LHC.  The resonant structure in the figure occurs at
$M_H\sim 2 M_T$ and is due to the fact that the amplitude for
$gg \rightarrow H$ gets an imaginary part for $M_H>2 M_T$,
as can be seen from the integral of Eq. \ref{intdef}.
The planned luminosity at the LHC is ${\cal L}=10^{34}/cm^2/sec$.
Hence a cross section of
 $1~pb$ corresponds to roughly $10^5$ events per year,
(a theorists year is typically taken to be $10^7 sec/year$).
In Section 3.3 we will investigate whether these $10^{5}$ events
are actually observable.
\subsection{Low Energy Theorem}

 A striking feature of our result for Higgs boson production
from gluon fusion is that it is independent of the
heavy quark mass for a light Higgs boson.  In fact
Eq.~\ref{heavym} can be derived
from the effective vertex,\cite{dawsp,keith,russ}
\beqn
{\cal L}_{\rm eff}&=&{\alpha_s \over 12 \pi} G_{\mu\nu}^A G^{A~\mu\nu}
\biggl({H\over v}\biggr)\nonumber \\
&=&{\beta_F \over g_s} G_{\mu\nu}^A G^{A~\mu\nu}
\biggl({H\over 2v}\biggr)(1-\delta),\nonumber
\label{effth}
\eeqn
where
\beq
\beta_F={g_s^3 N_H\over 24 \pi^2}
\eeq
 is the contribution of heavy fermion loops
to the $SU(3)$ beta function
and $\delta=2 \alpha_s/\pi$.\footnote{The $(1-\delta)$ term arises from
a subtlety in the use of the low energy theorem.  Since the Higgs
coupling to the heavy fermions is $M_f(1+{H\over v}){\overline f } f$,
the counterterm for the Higgs Yukawa coupling is fixed in terms of
the renormalization of the fermion mass and wavefunction.  The
beta function, on the other hand, is evaluated at $q^2=0$.
The $1-\delta$ term corrects for this mismatch.\cite{brat}}
($N_H$ is the number of heavy fermions
with $m>> M_H$.)
The effective Lagrangian of Eq.~\ref{effth}~ gives $ggH$ and
$gggH$ vertices and can be used to compute the
radiative corrections  of ${\cal O}(\alpha_s^3)$
 to gluon production.\cite{dawsp}  The correction in principle involve
2-loop diagrams.  However, using the effective vertices from
Eq. \ref{effth}, the ${\cal O}(\alpha_s^3)$ corrections can be found
from a 1-loop calculation.

In Fig.13 we show the radiatively corrected result for Higgs production from
gluon fusion.
Several important facts can be seen from this figure.  The first is that
there is very little dependance on the top quark mass and hence the
heavy  fermion mass limit (where all terms of ${\cal O}(M_H^2/M_T^2)$ are
neglected) is quite accurate.  When computing the lowest order result
from the triangle diagram of Fig. 8,
such as that shown in Fig. 11, it is ambiguous whether to use the one
or two loop equation for $\alpha_s$ and which structure functions to use,
a set fit to data using only  the lowest order (in $\alpha_s$)
 predictions or a set which includes
some higher order effects.\cite{mt}  The difference between the equations for
$\alpha_s$ and the different structure functions is
${\cal O}(\alpha_s^2)$
and hence higher order  in $\alpha_s$
when one is computing the ``lowest order"
result.  In Fig.13 we show 2 different definitions of the lowest order
result and see that they differ significantly from each other.  It
is interesting that the consistent ${\cal O}(\alpha_s^2)$ result (that
with one loop $\alpha_s$ and lowest order structure functions) is
closest to the radiatively corrected result.  We see that the radiative
corrections are large and
increase the production rate by about a factor of  $1.5$
from the lowest order result.

\subsection{Finding the Higgs Boson at the LHC}

We turn now to a discussion of search techniques for the
Higgs boson at the LHC.  For $M_H< 800~GeV$, gluon
fusion is the primary production mechanism, (for $M_T\sim 170~GeV$).
  At the present time, there are two
large detectors planned for the LHC; the ATLAS detector\cite{atlas}
and the CMS detector\cite{cms}.  We will present several
results from preliminary studies of these collaborations on the capabilities
of the LHC to discover the Higgs boson in various decay channels.
Detailed discussions of the experimental problems involved can
be found in the collaborations reports.

We have seen that the production rate for the Higgs boson at the LHC
is significant, $\sigma_H\sim .1 - 10~pb$ for $200~GeV < M_H
< 1~TeV$.  However, in order to see the Higgs boson it must decay into
some channel where it is not overwhelmed by the background.
For $M_H < 2 M_W$ the Higgs boson decays predominantly to
$b {\overline b}$ pairs, (remember that the Higgs coupling to fermions
is proportional to the fermion mass).  Unfortunately, the QCD production
of $b$ quarks is many orders of magnitude larger than Higgs production
and so this channel is thought to be useless.\cite{agel}
  One is led to consider
rare decay modes of the Higgs boson where the background may be smaller.
The decay channels which have received the most attention
are $H\rightarrow \gamma\gamma$ and
$H\rightarrow Z Z^*$.\cite{gg}\footnote{References to the many
studies of  the decays
$H\rightarrow\gamma\gamma$ and $H\rightarrow Z Z^*$
can be found in Refs. \cite{atlas,cms}.}
The branching ratios for these decays are shown in Fig. 14 and can
be seen to be  quite small.  (The rate for
off-shell gauge bosons,
 $H\rightarrow V V^*$, ($V=W^\pm,Z$)
must be multiplied by the relevant
branching ratio, $V\rightarrow f^{\prime} {\overline f}$.)

The  $H\rightarrow Z Z^*$ decay mode can lead to a final state
with $4$ leptons, $2$ of whose mass reconstructs to $M_Z$ while the
invariant mass of the $4$ lepton system reconstructs to $M_H$.  The
largest background to this decay is $t \overline{t}$ production
with $t\rightarrow W b \rightarrow (l \nu) ( c l \nu)$.  There are also
backgrounds from $Z b \overline{b}$ production, $Z Z^{*}$ production,
etc.
For $M_H=150~GeV$, the  ATLAS collaboration
estimates that there will be 184 signal events and 840 background
events in their detector in one year from $H\rightarrow Z Z^*
\rightarrow (4l)$ with the 4-lepton invariant mass in a
mass bin within $\pm 2 \sigma$ of $M_H$.\cite{atlas}
  The leptons from Higgs decay
tend to be more isolated from other particles than those coming
from the backgrounds and a series of isolation cuts
can be used to reduce the rate
to 92 signal and 38 background events.
 The ATLAS collaboration
claims that they will be able to discover the Higgs boson in the
$H\rightarrow Z Z^*\rightarrow l^+l^-l^+l^-$ mode for
$130~GeV < M_H < 180~GeV$ with an integrated luminosity
of $10^5 pb^{-1}$ (one year of running at the LHC
at design luminosity) and using both the electron and muon
signatures.  For $M_H< 130~GeV$, there are not
enough events since the branching ratio is too small (see Fig. 14),
while for $M_H> 180~GeV$ the Higgs search will proceed via the
$H\rightarrow ZZ$ channel, which we discuss in Section 4.3.

For $M_H< 130~GeV$, the Higgs boson can be searched for through
its decay to $2$ photons.  The branching ratio in this region
is about $10^{-3}$, so for a Higgs boson with $M_H\sim 100~GeV$
there will be about $3000$ events per year, ($\sigma\sim 30~ pb$ and
the LHC design luminosity is $10^{34}/cm^2/sec$.)  The
Higgs boson decay into the $\gamma \gamma$ channel
is an extremely narrow resonance in this region  with
a width around $1~KeV$.  From Fig. 14 we see that the branching
ratio for $H\rightarrow \gamma\gamma$ falls off rapidly with
increasing $M_H$ and so this decay mode is probably only useful in
the region $80~GeV < M_H < 130~GeV$.\footnote{The $H\rightarrow
\gamma\gamma$ branching ratio is sensitive to the top quark mass.
However, unlike the case $gg\rightarrow H$, there are
Feynman diagrams with $W$ bosons
in the loop which dominate over the top quark contribution.}

The irreducible background to $H\rightarrow \gamma\gamma$ comes from $q
\overline{q}\rightarrow \gamma\gamma$ and $gg \rightarrow \gamma
\gamma$ and is shown in Fig. 15.
In Fig. 16 we show the signal and the background for a Higgs boson
of mass $M_H=110~GeV$ at the LHC using the ATLAS detector.  Extracting
such a narrow signal from the immense background poses a formidable
experimental challenge.  The detector must have a mass resolution
on the order of $\delta m/m\sim 1.5 \%$ in order to be able to hope
to observe this signal.
For $M_H=110~GeV$ there are $1430$ signal events and $25,000$ background
events in a mass bin equal to the Higgs width.  This leads to a
ratio ,
\beq
 { {\rm Signal} \over \sqrt{{\rm Background}}}\sim 9
.\eeq
  A ratio greater than
$5$ is usually ${\it defined}$ as a discovery.  ATLAS  claims that
they will be able to discover the Higgs boson in this channel for
$100~GeV < M_H < 130~GeV$.  (Below $100~GeV$ the background is too
large and above $130~GeV$ the event rate is too small.)

There are many additional difficult experimental problems associated with
the decay channel $H\rightarrow \gamma \gamma$.  The most significant
of these is the confusion of a photon with a jet.  Since the cross
section for producing jets is so much larger than that of $H\rightarrow
\gamma\gamma$ the experiment must not mistake a photon for a jet more
than one time in $10^4$.  It has not yet been demonstrated that this
is experimentally feasible.

One might think that the decay $H\rightarrow \tau^+\tau^-$ would be useful
since as shown in Fig. 14
 its branching ratio is considerably larger than $H\rightarrow
Z Z^* $ and $H\rightarrow \gamma \gamma$, $BR(H\rightarrow \tau^+
\tau^-)\sim 3.5\%$ for $M_H=110~GeV$.  The problem is that for the
dominant production mechanism, $gg\rightarrow H$, the Higgs boson
has no transverse momentum and so the $\tau^+\tau^-$ invariant mass
cannot be reconstructed.  If we use the production mechanism,
$g g \rightarrow H g$, then the Higgs is produced at large transverse
momentum and it is  possible to reconstruct the $\tau\tau$ invariant
mass.  Unfortunately, however, the background from
$ q\overline{q}
\rightarrow \tau^+\tau^-$ and from $t \overline{t}$ decays overwhelms
the signal.\cite{keith}

\subsection{Higgs Boson Production at the Tevatron}

Since it will be some years before the LHC comes into operation
it is worth considering whether any relevant limits on the Higgs
boson can be obtained from the existing hadron collider, the
Tevatron, which is a $p \overline{p}$ collider with an energy
$\sqrt{s}=1.8~TeV$.  For a Higgs boson mass of $60~GeV$  the production
 cross
section at the Tevatron is roughly $4~pb$.  At a luminosity
of ${\cal L}\sim 10^{31}/cm^2/sec$ this yields $400$ Higgs events
per year.  To look at these events in the $\gamma\gamma$ decay
mode we must multiply by a branching ratio of $10^{-3}$ which
leaves $.4$ Higgs events per year.  Even with the main injector,
which will increase the luminosity by about a factor of $10$,
finding the Higgs boson through this decay channel is clearly
hopeless at the Tevatron.

Recently it has been suggested that it may be fruitful to look for
the Higgs boson at the Tevatron through the production mechanism
$q \overline{q}^{\prime}\rightarrow W H$,
 shown in Fig. 17.\cite{stange,gunhan}
  The various production
mechanisms which are relevant for producing a Higgs boson at the
Tevatron are shown in Fig. 18.
The cross section for
 $q \overline{q}^{\prime}
\rightarrow W H$  is about $1~pb$ which gives on the
order of $20$ events/year if we look at the decays $W\rightarrow
e{\overline \nu}$ and $W\rightarrow \mu{\overline \nu}$.
  Unfortunately, the cuts to eliminate
backgrounds make this mechanism not viable.
Potential upgrades to increase either the luminosity or the energy
at the Tevatron may make this a viable option.\cite{stange,gunhan}
The $WH$ production mechanism may also be useful at the LHC to look
for a Higgs boson in the $M_H\sim 100~GeV$ region.

\section{Higgs Boson Production from Vector Bosons}
\subsection{The Effective $W$ Approximation}

We turn now to the  study  of
the couplings of the Higgs bosons to gauge
bosons.  We begin by studying the diagram in Fig. 19.
Naively, one expects this diagram to give a negligible contribution
to Higgs production because of the two $W$ boson propagators.  However,
it  turns out that this production mechanism
can give an important contribution.
The diagram of Fig. 19 can be interpreted in parton model language
as the resonant scattering of two $W$ bosons to form a Higgs
boson\cite{effw} and we
can compute the distribution of $W$ bosons in a quark in an
analogous manner to the computation of the distribution of photons
in an electron.\cite{effgam}
By considering the $W$ and $Z$ gauge bosons as partons, calculations
involving gauge bosons in the intermediate states can be considerably
simplified.

We define orthogonal
polarization tensors for a $W$ boson with momentum
$k=(k_0,0,0,\mid {\vec k}\mid)$:
\beqn
{\rm Transverse}:\quad \epsilon_\pm&=& {1\over \sqrt{2}}
(0,1,\pm i, 0) \nonumber \\
{\rm Longitudinal}:\quad \epsilon_L&=&
{1\over M_W}(\mid {\vec k}\mid, 0,0,k_0).
\label{eps}
\eeqn
For large momentum, $k_0>>M_W$, we have
\beq
\epsilon_L \sim {k\over M_W}+{M_W\over 2 k_0}(-1,0,0,1).
\label{epsl}
\eeq
The first term in $\epsilon_L$ gives zero for the coupling of
longitudinal $W$'s to massless fermions and so the
longitudinal coupling is suppressed by $M_W/k_0$ relative
to the transverse coupling to massless fermions.
It is instructive to begin by
computing the coupling of a Higgs boson
to two longitudinal $W$ bosons, (Fig. 20).
The amplitude is given by
\beq
{\cal A}(H\rightarrow W^+_LW^-_L)
=g M_W \epsilon_L(p_+)\cdot\epsilon_L(p_-).
\eeq
Using Eq.~\ref{eps}  we have for $M_H>>M_W$,
\beq
{\cal A}(H\rightarrow W^+_LW^-_L)
={g M_H^2\over 2 M_W}.
\eeq
The longitudinal coupling of the Higgs boson to $W$ bosons is
enhanced for heavy Higgs bosons!  It is this fact
 which makes
the process of Fig. 19 relevant.

In order to treat the $W^\pm$ and $Z$ bosons as partons, we consider them as
on-shell physical bosons.  We make the approximation that the
partons have zero
transverse momentum, which ensures that the longitudinal and transverse
projections of the $W$ and $Z$ partons are uniquely specified.
We want to be able to write a parton level relation ship:
\beq
\sigma(q_1+q_2\rightarrow q_1^\prime+X) =
\int^1_{M_W\over E} dx f_{q/W}(x)\sigma(W+q_2\rightarrow X).
\label{pardef}
\eeq
The function $f_{q/W}(x)$ is  called the distribution of
$W$'s in a quark and it is defined by Eq. \ref{pardef}.

The amplitude for a $W$ with polarization vector $\epsilon_i$
 to scatter from
a quark $q_2$ to the final state $X$ (see Fig. 21) is:
\beq
i{\cal A}_i(W+q_2\rightarrow X)=\epsilon_i\cdot {\cal M} \sqrt{E_q},
\eeq
where $E_q$ is the quark energy and we have replaced the
$W-q_2-X$ vertex
by an effective coupling ${\cal M}_\mu\sqrt{E_q}$.  Averaging over the
quark spin and dropping terms suppressed by $M_W^2/E^2$, we find
\beq
d\sigma(W_i+q_2\rightarrow X)={1\over 8 k_0}\mid \epsilon_i
\cdot{\cal M}\mid^2 d \Gamma_X,
\eeq
where the Lorentz invariant phase space of the final state $X$ is
$d\Gamma_X$.

We now consider the two body scattering process of Fig. 22 which
gives the amplitude,
\beq
i{\cal A}_i(q_1+q_2\rightarrow q_1^\prime+X)
={g\over 2\sqrt{2}}{\overline u}(p^\prime)\epsilon_i^* (1-\gamma_5)
u(p)\epsilon\cdot{\cal M} {\sqrt{E_q}\over k^2-M_W^2}.
\eeq
Because of the factor $1/(k^2-M_W^2)$, the cross section is dominated
by small angles since
\beq
k^2\sim E^2 (1-x)\theta^2,
\eeq
where $\theta$ is the angle between the $W$ and the outgoing quark,
$q_1^\prime$.

The spin averaged total cross section is then
\beq
\sigma_i(q_1+q_2\rightarrow q_1^\prime+X)
={1\over 32 E E_q}\int {d^3 p^\prime\over (2 \pi)^3}
{\mid {\cal A}_i(q_1+q_2\rightarrow q_1^\prime+X)\mid^2
\over E^\prime} d \Gamma_X .
\label{sigg}
\eeq
The effective $W$ approximation consists of replacing the current,
$\mid \epsilon_i\cdot {\cal M}\mid$,
 and the phase space, $d\Gamma_X$,
by their values  when $k^2 \rightarrow M_W^2$ and the $W$
is emitted in the forward direction, $\theta\rightarrow 0$.
Using the definition of Eq.~\ref{pardef} and the polarizations
of Eq. ~\ref{eps}
we find the  $W$ distributions in a quark from Eq.
\ref{sigg},\cite{effw}
\beqn
f_{q/W}^L(x)&=&{g^2\over 8 \pi^2}{M_W^2\over E^2}
{1\over x}\int {\theta d\theta\over (\theta^2+{M_W^2\over E^2(1-x)})^2}
\nonumber \\
&=&{g^2\over 16 \pi^2}{1-x\over x}\nonumber \\
f_{q/W}^T(x)&=&{g^2\over 64\pi^2 x}\log\biggl({4 E^2\over M_W^2}\biggr)
\biggl[1+(1-x)^2\biggr],
\label{wdists}
\eeqn
where we have averaged over the $2$ transverse polarizations.
The logarithm in Eq.~\ref{wdists} is the same logarithm which appears in the
effective photon approximation.\cite{effgam}
The result of Eq. \ref{wdists} violates our intuition that longitudinal gauge
bosons don't couple to massless fermions.
  However, the integral
over $d \theta$ picks out the $\theta\rightarrow 0$ region and hence the
subleading term in the polarization tensor of Eq. \ref{epsl}.

It is now straightforward to compute
the rates for
 processes involving $WW$ scattering.
The  hadronic cross section can be written in terms of a luminosity
of $W$'s in the proton,
\beq
\sigma_{pp\rightarrow WW\rightarrow X}(s)=\int_{\tau_{min}}^1
d \tau {d {\cal L}\over d \tau}\mid_{pp/WW}\sigma_{WW\rightarrow X}
(\tau s)
\eeq
where the luminosities are defined:
\beqn
 {d {\cal L}\over d \tau}\mid_{pp/WW}&=&\sum_{ij}
\int_\tau^1{d \tau^\prime \over \tau^\prime}
\int^1_{\tau^\prime}{dx\over x}
f_i(x)f_j\biggl({\tau^\prime\over x}\biggr)
 {d {\cal L}\over d \zeta}\mid_{q_i q_j/WW}
\nonumber \\
 {d {\cal L}\over d \tau}\mid_{qq/WW}&=&
\int_\tau^1{dx\over x}
f_{q/W}(x)f_{q/W}\biggl({\tau\over x}\biggr).
\label{lumo}
\eeqn
($f_i(x)$ are the quark distribution functions in the proton and
$\zeta\equiv \tau/\tau^{\prime}$).
Of course this entire derivation can also be performed for $Z$ bosons.
In addition, the luminosities of Eq. \ref{lumo} can be trivially adapted
to find the distribution of gauge bosons in the
electron.\cite{dawros}

In Fig. 23 we  show the luminosity of transverse and longitudinal
gauge bosons in the proton at the LHC.  It is interesting to
note that the transverse luminosity is several orders of magnitude
larger than the longitudinal luminosity due to the
enhancement from the logarithm in $f_{q/W}^T$,
as can be seen from Eq. \ref{wdists} .  However, because the coupling
of longitudinal gauge bosons  to a heavy  Higgs boson and
to heavy fermions\footnote{
For heavy fermions, the ${\overline \psi} \psi W_L$ coupling
is proportional to $M_f/M_W$.} is enhanced the dominant
contribution to a physical process is often from longitudinal
gauge boson scattering.

The effective $W$ approximation is particularly useful in models
where the electroweak symmetry breaking is  due not to the Higgs
mechanism, but rather to some strong interaction dynamics (such
as technicolor models).  In these models one typically estimates
the strengths of the $3$ and $4$ gauge boson couplings due to the
new physics.  These interactions can then be folded into the
luminosity of gauge bosons in the proton (or the electron) to
get estimates of the size of the new physics effects.

\subsection{Does the Effective $W$ approximation Work?}

It is interesting to ask if QCD effects spoil the effective $W$
approximation.\cite{sopkun}  Diagrams of the sort shown in Fig. 24
for example cannot be calculated
 within the context of the effective
$W$ approximation.
The diagram of Fig. 24a gives a contribution to
the cross section of ${\cal O}(\alpha_s)$
which is proportional to $Tr(T_A)=0$.  Most of the remaining
contributions to the QCD corrections (Figs. 24b and c) can be
absorbed in the definition of the structure functions to next
to leading order.  It has been demonstrated by explicit
calculation that the QCD corrections to the effective $W$ approximation
are small, of order $10\%$.\cite{qcdw}

We turn now to a discussion of Higgs boson production from vector
boson fusion and compare results obtained with and without
the effective $W$ approximation.
In the effective $W$ approximation,
\beq
\sigma(pp\rightarrow H)=
{16 \pi^2\over M_H^3}\Gamma(H\rightarrow W^+W^-)
\tau {d {\cal L}\over d \tau}\mid_{pp/WW}.
\eeq
A heavy Higgs boson will
quickly decay into $W^+W^-$ pairs with a width
\beq
\Gamma(H\rightarrow W^+W^-)\sim 2 \Gamma(H\rightarrow ZZ)
\sim {G_F M_H^3\over 8 \sqrt{2}\pi}.
\eeq
As the Higgs mass approaches a TeV, its mass becomes comparable
to its width.  A useful form to remember is that
summed over $W^{\pm}, Z$,
\beq
\Gamma(H\rightarrow VV)\sim {M_H^3\over 2} \quad {\rm (TeV~ units)}
{}.
\eeq
Therefore we cannot consider simply Higgs production for a heavy
Higgs boson, but must
consider the $W^+W^-$ or $Z Z$ final state where the Higgs
boson
contributes to an $s$- channel resonance.
All of the  diagrams contributing to $VV\rightarrow VV$ scattering
must be included in order to obtain a gauge invariant result.

There is an extensive literature demonstrating the validity of
the effective $W$ approximation at the SSC and we discuss the
relevant physics  points here.\footnote{Once the validity of the effective
$W$ approximation was established for the SSC, theorists
didn't bother to redo the calculations for the LHC!}
In Fig. 25,  we show the contribution of $W^+W^-$ and $ZZ$
scattering
to the final state $ZZ$
for both longitudinal and transverse gauge boson intermediate states
 through the
mechanism of Fig. 19 .
The Feynman diagrams contributing to the process
$W^+W^-\rightarrow ZZ$ are shown in Fig. 26.   A similar set of diagrams
contributes to $ZZ\rightarrow ZZ$.
It is clear that except for $M_{ZZ}$ near the Higgs
pole, it is a poor approximation to keep only the longitudinal modes.
Both transverse and longitudinal gauge bosons contribute to the
physical amplitude.

We can also investigate whether it makes sense to include only
the $s$-channel Higgs exchange diagram.
 From Fig. 27
we can see that this
is a
poor approximation except for $M_{WW}$ near the Higgs boson mass.
This figure also includes the contribution to $W^+W^-$ production
from the direct scattering $q \overline{q}\rightarrow W^+W^-$. This
process is often called the background to Higgs production and
is much larger than the
contribution from $W^+W^-$ scattering.  As the Higgs boson becomes
increasingly massive, its width becomes wider and the tiny Higgs
bump shown in Fig. 27 becomes impossible to observe.

Higgs boson production through vector boson fusion
 can of course be computed numerically
without the use of the effective $W$ approximation.
In Fig.~28 we show a comparison of the exact numerical calculation
for the process $pp\rightarrow WW\rightarrow ZZ$ compared with
that derived from the effective $W$ approximation.
The agreement is excellent
and the effective $W$ approximation is accurate to within a
factor of two even far from the Higgs pole.\cite{abel,gun}
  This did not have
to be the case since
the effective $W$ approximation
does not contain diagrams where the $W$ is radiated off the
incoming or outgoing quark lines (see Fig. 29).
We conclude
from this series of plots, (Figs. 25, 27 and 28),
 that the effective $W$ approximation can be used with
confidence even away from the Higgs pole if both transverse and
longitudinal gauge boson contributions and if  all scattering diagrams
(not just $s-$ channel Higgs exchange) are included.
The most important use of the effective $W$ approximation is
the study of very massive, strongly interacting Higgs bosons, which we
consider in Section 5.

\subsection{Searching for a Heavy Higgs Boson at the LHC}

We now have the tools necessary to discuss the search for a
very massive Higgs boson.   The various production mechanisms
contributing to Higgs production  at the LHC are shown in Fig. 30.
For $M_H< 800~GeV$ the dominant production mechanism at the LHC is gluon
fusion, as discussed in  Section 3.1.  For heavier Higgs masses,
the $WW$ fusion mechanism becomes important.
Other mechanisms, such as $gg\rightarrow t {\overline t} H$,
are quite small at the LHC.\cite{tth}  It is obvious
from Fig. 30 that searching for a Higgs boson on the TeV mass scale
will be extremely difficult.  For example, a $700~GeV$ Higgs boson
has a cross section near $1~pb$ leading to around $10^5$ events/LHC
year.  The cleanest way to see these events is the so-called
``gold-plated" decay channel,
\beq
H\rightarrow ZZ\rightarrow l^+l^- l^+ l^- .
\eeq
The lepton pairs will reconstruct to the $Z$ mass and the $4$
lepton invariant mass will give the Higgs mass.
Since the branching ratio,
 $Z\rightarrow e^+e^- + \mu^+\mu^-$ is $\sim .06$
the number of events for a $700~ GeV$ Higgs is reduced to around
$360$ four lepton events per year.  Since this number will be further
reduced by cuts to separate the signal from the background, it
is clear that this channel will run out of events as the Higgs
mass becomes heavier.\cite{zzgold}

In Fig. 31, we show a Monte Carlo simulation of the capabilities
of the ATLAS detector at the LHC to observe a Higgs boson with mass
$M_H=800~GeV$ through the $4$ lepton decay channel.\cite{atlas}
  The upper curve
shows all of the events including the background from $ZZ$ continuum
production.  A series of kinematic cuts is applied until the lower
curve is reached, where the Higgs bump can be seen.  The ATLAS
collaboration claims that they will be able to discover the
Higgs boson in the mass region $130< M_H < 800~GeV$ in the $4$
lepton channel.  (Similar results are found by the CMS
collaboration.\cite{cms}).

In order to look for still heavier Higgs bosons, one can look in the
decay channel,
\beq
H\rightarrow ZZ\rightarrow l^+l^- \nu \overline{\nu}.
\eeq
Since the branching ratio, $Z\rightarrow \nu \overline{\nu}\sim 20~\%$,
this decay channel has  a larger rate than the
four lepton channel.  However, the price is that
because of the neutrinos, events of this type cannot be fully
reconstructed.
An example of this signal is shown in Fig. 32.  This channel extends
the Higgs mass reach of the LHC slightly.

Another idea which has been proposed is to use the fact that events
coming from $WW$ scattering have outgoing jets at small angles,
whereas the $WW$ background coming from $q \overline{q}\rightarrow
W^+W^-$ does not have such jets.\cite{barger} Additional sources of
background to Higgs detection
such as $W$+ jet production have jets at all angles.
It is not yet clear whether this idea will be useful in the
search for a heavy Higgs.

In this section and in Section 3.3, we have seen that
the LHC will have the capability
to observe the Higgs boson in the mass region from $100 < M_H
< 800~GeV$.
We now return to the Goldstone boson sector of the theory in an
attempt to learn something about the Higgs boson in the regime
where it is too heavy to be observed at the LHC.
\section{Strongly Interacting Higgs Bosons}

We can see from the Feynman rules of Fig. 4 that as the Higgs boson
becomes heavy, its self interactions become large and
it becomes
strongly interacting.  For $M_H>1.4~TeV$, the total Higgs boson
decay  width is
larger than its mass and it no longer makes sense to think of
the Higgs boson as a particle.
This regime can most easily be studied by going to the Goldstone
boson sector of the theory.  In Feynman gauge, the three Goldstone
bosons, $\omega^\pm,z$, have mass $M_{\omega,z}=M_{W,Z}$ and have
the interactions,\cite{cfh}
\beq
V={M_H^2\over 2v}H\biggl(H^2+Z^2+2 \omega^+\omega^-\biggr)
+{M_H^2\over 8 v^2}\biggl(H^2+z^2+2\omega^+\omega^-\biggr)^2.
\label{vgold}
\eeq
The Feynman rules corresponding to Eq. \ref{vgold} are given
in Fig. 33.

Calculations involving only the Higgs boson and the  Goldstone
bosons are easy since they involve only scalars. For example
the amplitude for $\omega^+\omega^-\rightarrow \omega^+\omega^-$\cite{lqt}
 can
be found from the Feynman diagrams of Fig. 34:
\beq
{\cal A}( \omega^+\omega^-\rightarrow
\omega^+\omega^-)=-{M_H^2\over v^2}
\biggl({s\over s-M_H^2}+{t\over t-M_H^2}\biggr),
\label{wwwwgold}
\eeq
where $s,t,u$ are the Mandelstam variables in the $\omega^+\omega^-$ center
of mass frame.
It is instructive to compare Eq. \ref{wwwwgold}
with what
is obtained
 by computing $W^+_LW^-_L\rightarrow W^+_L W^-_L$
using real  longitudinally polarized
gauge bosons and extracting the leading
power of $s$ from each diagram\cite{duncan}:
\beqn
 {\cal A}(W^+_LW^-_L\rightarrow W^+_LW^-_L)&=&
-{1\over v^2}\biggl\{
-s-t+2 M_Z^2 +{2 t\over s}\biggl(M_Z^2-4 M_W^2\biggr)
+{2 M_Z^2 s\over t-M_Z^2}
\nonumber \\
&&-8 \sin^2\theta_W M_W^2 \biggl({M_Z^2 s\over t(t-M_Z^2)}
\biggr)+{s^2\over s-M_H^2}+{t^2\over t-M_H^2}\biggr\}.\nonumber \\
&&
\label{wwwwreal}
\eeqn
{}From Eqs. ~\ref{wwwwgold} and ~\ref{wwwwreal} we find an
amazing result,
\beq
 {\cal A}(W^+_LW^-_L\rightarrow W^+_LW^-_L)=
{\cal A}( \omega^+\omega^-\rightarrow
\omega^+\omega^-)+{\cal O}\biggl({M_W^2\over s}\biggr).
\eeq
This result means that instead of doing the complicated
calculation with real gauge bosons, we can instead do
the easy calculation with only scalars if we are at an
energy far above the $W$ mass and are interested only
in those effects which are enhanced by $M_H^2$.
(The interactions of the transverse gauge bosons are ${\cal O}(g^2)$
and have no $M_H^2$ enhancement.)
  This is a general
result and has been given the name of the electroweak
equivalence theorem.\cite{et}

The formal statement of the electroweak equivalence
theorem is that
\beq
{\cal A}(V_L^1 V_L^2....V_L^N\rightarrow
V_L^1 V_L^2....V_L^{N^\prime})=(i)^N(-i)^{N^\prime}
{\cal A}(\omega_1\omega_2...\omega_N\rightarrow
\omega_1\omega_2...\omega_{N^\prime})
+{\cal O}\biggl({M_V^2\over s}\biggr),
\label{equivth}
\eeq
where $\omega_i$ is the Goldstone boson corresponding to
the longitudinal gauge boson, $V_L^i$.  In other words, when
calculating scattering amplitudes of longitudinal
gauge bosons  at high energy, we can
replace the ${\it external}$ longitudinal gauge bosons
by Goldstone bosons.  A formal proof of this theorem
can be found in Ref. \cite{et}.

The electroweak equivalence theorem is extremely useful in
a number of  applications.  For example, to compute the
radiative corrections to $H\rightarrow W^+W^-$\cite{scott}
or to $W^+_LW^-_L\rightarrow W^+_LW^-_L$\cite{sdsw}, the dominant
contributions which are enhanced by $M_H^2/M_W^2$ can be
found by computing the one loop corrections to
$H\rightarrow \omega^+\omega^-$ and to $\omega^+
\omega^-\rightarrow \omega^+\omega^-$ which involve only scalar
particles.
Probably the most powerful application
of the electroweak equivalence
theorem is, however, in the search for the physical effects of strongly
interacting gauge bosons which we turn to now.

\subsection{$M_H\rightarrow\infty$, The Non-Linear Theory}

So far we have considered searching for the Higgs boson in
various mass regimes.  In this section we will consider the consequences
of taking the Higgs boson mass much
heavier than the energy scale being probed\cite{abl}.
In fact, we will take the limit $M_H\rightarrow \infty$ and
assume that the effective Lagrangian for electroweak
symmetry breaking is determined by new physics outside the reach
of future accelerators such as the LHC.  Since we do not know
the full theory, we must build the effective Lagrangian out of
all operators consistent with the unbroken symmetries.  In particular,
we must include operators of all dimensions, whether or not
they are renormalizable.  In this way we construct the most general
effective Lagrangian that describes electroweak symmetry breaking.

To specify the effective Lagrangian, we must first fix the pattern
of symmetry breaking.  We will assume that the global symmetry in
the scalar sector of the model is
 $SU(2)_L \times SU(2)_R$ as in the minimal Standard Model, (see
for example, Eq. \ref{vgold}).
In this case, the Goldstone bosons can be described in terms
of the field\cite{leut}
\beq
\Sigma\equiv e^{{i \omega\cdot\tau\over v}}.
\eeq
This is reminiscent of the Abelian Higgs model
where we took,
\beq \Phi={1\over \sqrt{2}}e^{{i\chi\over v}}(H+v).
\eeq
Under the global symmetry the $\Sigma$ field
transforms as,
\beq
\Sigma \rightarrow L^\dagger \Sigma R.
\label{trans}
\eeq
It is straightforward to write down the most general $SU(2)_L\times
U(1)_Y$ gauge invariant Lagrangian which
respects the global symmetry of Eq. \ref{trans} and has no more
than $2$ derivatives acting on the $\Sigma$ field,
\beq
{\cal L}={v^2\over 4}D_\mu\Sigma^\dagger D^\mu\Sigma
-{1\over 2}Tr \biggl({\hat W}^{\mu\nu}{\hat W}_{\mu\nu}\biggr)
-{1\over 2}Tr \biggl({\hat B}_{\mu\nu} {\hat B}^{\mu\nu}\biggr),
\label{chiral}
\eeq
where the covariant derivative is given by,
\beq
D_\mu\Sigma =\partial_\mu \Sigma +{i\over 2} g
{\hat W}_\mu^i\tau^i \Sigma-{i\over 2} g ^{\prime}
 {\hat B}_\mu \Sigma \tau_3.
\eeq
The gauge field kinetic energies are now matrices:
\beqn
{\hat W}_{\mu\nu}& \equiv &{1\over 2}\biggl(
\partial_\nu W_\mu-\partial_\mu W_\nu
-{i\over 2} g [ W_\mu,W_\nu]\biggr)\nonumber \\
{\hat B}_{\mu\nu}&=& {1\over 2}
 \biggl(\partial_\nu B_\mu-\partial_\mu B_\nu
\biggr)\tau_3
\eeqn
with $W_\nu\equiv W_\nu^i \cdot \tau_i$.
We will also assume a custodial $SU(2)_C$ symmetry.\cite{cgg}  This is the
symmetry which forces $\rho=M_W^2/(M_Z^2\cos\theta_W)=1$.
The pattern of global symmetry breaking is then $SU(2)_L\times
SU(2)_R \rightarrow SU(2)_C$.
In unitary gauge, $\Sigma=1$ and it is easy to see that Eq. \ref{chiral}
generates mass terms for the $W$ and $Z$ gauge bosons.
The Lagrangian of
 Eq. \ref{chiral}
{\bf is} the Standard Model with $M_H\rightarrow \infty$.

Using the Lagrangian of Eq. \ref{chiral}
 it is straightforward to compute
Goldstone boson scattering
 amplitudes such
as\cite{wb}
\beq
{\cal A}(\omega^+\omega^-\rightarrow zz)={s\over v^2}\equiv A(s,t,u)
\label{pipi}
\eeq
which of course agree with those found in the Standard Model
when we take $M_H^2 >> s$.  Because of the custodial $SU(2)_C$ symmetry,
the various scattering amplitudes are related:
\beqn
{\cal A}(\omega^+ z\rightarrow \omega^+ z)&=& A(t,s,u)\nonumber \\
{\cal A}(\omega^+ \omega^-\rightarrow \omega^+
\omega^-)&=& A(s,t,u)+A(t,s,u)\nonumber \\
{\cal A}(\omega^+ \omega^+\rightarrow \omega^+
\omega^+)&=& A(t,s,u)+A(u,t,s)\nonumber \\
{\cal A}( z z\rightarrow z z)&=& A(s,t,u)+A(t,s,u)+A(u,s,t).
\label{isorel}
\eeqn
Using the electroweak equivalence theorem, the Goldstone boson scattering
amplitudes can be related to the amplitudes for longitudinal gauge boson
scattering.  The effective $W$ approximation can then be
used to find the physical scattering cross sections for hadronic
and $e^+e^-$ interactions.

The relationships of Eq. \ref{isorel}
 were discovered by Weinberg\cite{wb}
 over 20 years ago
for the case of $\pi-\pi$ scattering.\footnote{There is an exact
analogy between $\pi\pi$ scattering and $W^+_LW^-_L$ scattering
with the replacement $f_\pi\rightarrow v$.}
Amplitudes which grow with $s$ are
a disaster
for perturbation theory
since eventually they will violate perturbative
unitarity as we will discuss in Sec. 6.1.  Of course, this simply tells
us that there must be some new physics at high energy.

Eq. ~\ref{chiral} is a non-renormalizable effective
 Lagrangian which
must be interpreted as an expansion in powers of $s/\Lambda^2$,
where $\Lambda$ can be taken to be the scale of new physics (say
$M_H$ in a theory with a Higgs boson).  At each order in the
energy expansion new terms will be generated which will cancel
the singularities generated by the order below.
To ${\cal O}(s^2)$, the infinities which arise at one loop can
all be absorbed by defining renormalized parameters, $L_i(\mu)$.
The coefficients thus depend on the renormalization scale $\mu$.
  At ${\cal O}(s^2/
\Lambda^4)$ we have the interaction terms,
\beqn
{\cal L}&=&{L_1\over 16\pi^2}\biggl[Tr\biggl(D^\mu\Sigma^\dagger
D_\mu\Sigma\biggr)\biggr]^2 +
{L_2\over 16\pi^2}\biggl[Tr\biggl(D^\mu\Sigma^\dagger
D_\nu\Sigma\biggr)\biggr]^2
\nonumber \\
&&-{i g L_{9L}\over 16\pi^2}Tr\biggl({\hat W}^{\mu\nu}D_\mu\Sigma
D_\nu\Sigma^\dagger\biggr)
-{i g^{\prime}L_{9R}\over 16\pi^2}Tr\biggl({\hat B}^{\mu\nu}
D_\mu\Sigma^\dagger
D_\nu\Sigma\biggr)\nonumber
\\
&&+{g g^\prime L_{10}\over 16 \pi^2} Tr \biggl(\Sigma
{\hat B}^{\mu\nu}\Sigma^{\dagger}{\hat W}_{\mu\nu}\biggr)
.\label{fourl}
\eeqn
This is the most general
$SU(2)_L\times U(1)_Y$ gauge invariant
set of interactions of ${\cal O}(1/\Lambda^4)$
which preserves the custodial $SU(2)_C$.  The coefficients,
$L_i$ have information about the underlying dynamics of the
theory.  By measuring the various coefficients one might hope
to learn something about the mechanism of electroweak symmetry
breaking even if the energy of an experiment is below the scale at
which the new physics occurs.

The $L_{10}$ interaction contributes to non-Standard Model
2- and 3-gauge boson interactions.  New physics at LEP is often
parameterized in terms of the contributions to the gauge
boson 2-point functions, (the so-called
``oblique corrections'').  In a theory without a
custodial $SU(2)_C$ symmetry, there are three possible interactions,
often called S,T, and U.\cite{pt} Since we have assumed a custodial
$SU(2)_C$ symmetry, there is only one  interaction
contributing to non-Standard Model 2- gauge boson interactions and
we have
\beq L_{10}(M_Z)=-\pi S. \eeq
The $L_{10}$ interaction
contributes to $\gamma-Z$ mixing and is limited by precision
electroweak measurements at LEP:\cite{dvlep,langnew}
\beq
-1.1 < L_{10}(M_Z) < 1.8
\eeq
at the $90~\%$ confidence level from measurements of the
total $Z$ width.
The $L_{9L}, L_{9R}$ and $L_{10}$ coefficients all contribute
to $3$ gauge boson interactions, while the
$L_{1}, L_{2},L_{9L}$, and $ L_{9R}$  interactions
contribute to $4$ gauge
boson interactions.  Within this framework  the 2,3, and 4-point
 gauge boson interactions are related by the gauge invariance
of Eq. \ref{fourl}.

Effects of the new interactions can be looked for in $e^+e^-$ interactions,
as in Fig. 35.
At high energy, there is a delicate cancellation between $t$-
channel $\nu$ exchange and $s$- channel $\gamma$ and $Z$ exchange
in the process $e^+e^-\rightarrow W^+W^-$.\cite{buras}  If this cancellation
is spoiled there is a contribution to the cross section which
grows with energy.\cite{hol}  The Lagrangian
of Eq.~\ref{fourl} contributes
terms which grow with $s$ to the $3$  gauge boson vertices
shown in Fig. 35 which is potentially
measurable at LEPII.
Unfortunately, these effects tend to be rather small in all  models
which have been considered.

The effects of the non-standard model couplings of Eq.
\ref{fourl}
can also be searched for in hadron machines which are
sensitive to both the three and
 four gauge boson vertices.\cite{boud, bdv,fls,baghan}
To study strong interactions with the Lagrangian of Eq. ~\ref{fourl}
one must use the effective $W$ approximation to get results for $pp$
scattering.  This has the result that the calculation has a rather
limited region of validity,
\beq
M_W^2 < {\hat s}    < \Lambda^2.
\eeq
 In
Fig. 36, we show the effects  at the LHC of
turning on small values of the $L_i$ as compared to the lowest
order result  and see that the effects
 are quite small.\footnote{The $W^+W^+$ channel is
advantageous in the search for strongly interacting
symmetry breaking effects since there is no $q {\overline q}$
background.}
We define a signal as ``observable''  if it induces a change in
the integrated cross section of greater than $50\%$, which implies
that the LHC will be sensitive to $\mid L_i \mid > 1$. (More
precise values can be found in Ref. \cite{bdv,fls}.)

 \subsection{Coefficients of New Interactions in  a Strongly
Interacting Symmetry Breaking Sector}
It is instructive to estimate the size of the  $L_i$ coefficients in
typical theories.  Using the effective Lagrangian approach this can be
done in a consistent way.  We first consider a model in which
we couple the Goldstone bosons to a scalar, isoscalar resonance
like the Higgs boson.  We assume that the $L_i$ are
dominated by tree-level exchange of the scalar boson.  By integrating
out the scalar particle
 and matching coefficients at the scale $M_H$, we
find\cite{mod}
\beqn
L_1(\mu)&=&{64 \pi^3\over 3}{\Gamma_H v^4\over M_H^5}
+{1\over 24}\log\biggl({M_H^2\over \mu^2}\biggr)\nonumber \\
L_2(\mu)&=& L_{9L}(\mu)=L_{9R}(\mu)=-L_{10}(\mu)=
{1\over 12}\log\biggl({M_H^2\over \mu^2}\biggr),
\eeqn
where $\Gamma_H$ is the width of the scalar into Goldstone bosons.  If
we naively take
\beq
\Gamma_H={3 M_H^3\over 32 \pi v^2}
\eeq
as in the Standard Model, we find for $M_H=2~TeV$ and $\mu=1.5~TeV$,
$L_1=.33$ and $L_2=.01$.

We can also consider a second model for the $L_i$ and assume that the
coefficients are dominated by tree-level exchange of a $\rho$-like
particle with spin and isospin one.  Integrating out the $\rho$
and matching coefficients at the scale $M_\rho$ we find\cite{mod}
\beqn
L_1(\mu)&=&{1\over 24}\biggl[-{96 \pi^2 f^2\over M_\rho^2}+
\log\biggl({M_\rho^2\over \mu^2}\biggr)\biggr]\nonumber \\
L_2(\mu)&=&{1\over 12}\biggl[{48 \pi^2 f^2\over M_\rho^2}+
\log\biggl({M_\rho^2\over \mu^2}\biggr)\biggr]\nonumber \\
L_{9L}(\mu)=L_{9R}(\mu)&=&{1\over 12}\biggl[{96 \pi^2 f
F_\rho\over M_\rho^2}+
\log\biggl({M_\rho^2\over \mu^2}\biggr)\biggr]\nonumber \\
L_{10}(\mu)&=&-{1\over 12}\biggl[{48 \pi^2 F_\rho^2\over M_\rho^2}+
\log\biggl({M_\rho^2\over \mu^2}\biggr)\biggr]\nonumber \\
\eeqn
where the constant $f$ is related to the width $\Gamma_\rho$,
\beq
\Gamma_\rho={1\over 48 \pi}{f^2\over v^4} M_{\rho}^3
\eeq
and $F_\rho$ is defined by
\beq
\langle 0 \mid V_\mu^i \mid \rho^k(p)\rangle
=\delta^{ik} \epsilon_\mu F_\rho M_\rho.
\eeq
We can use large N scaling arguments to estimate $f$ and $F_\rho$.
For $M_\rho=2~TeV$ and $\mu=1.5~TeV$, we find $L_1=-.31$, $L_2=.38$,
$L_9=1.4$ and $L_{10}=-1.5$.
Since we estimated that the LHC will be sensitive to $\mid L_i\mid
< 1$, we see
that the LHC will indeed probe electroweak symmetry breaking in the TeV
region.

One can estimate the amount of time it would take to see a signal
of a strongly interacting electroweak symmetry breaking sector
at the LHC.
The signal has no resonance shape and will be a small excess of events
over that predicted from the lowest order Lagrangian of
Eq. \ref{chiral}.
  For example, a model with the $L_i$ having values  corresponding
to a $2.5~TeV$ techni-rho would only be observable at the LHC
with five years of running! (and then there would be only 14 signal
events in the optimal $W^+Z$ channel!).\cite{baghan}
  It is clear that this is an extremely
difficult way in which to look for evidence of the Higgs boson.

\section{Indirect Limits on the Higgs Boson Mass}

In the first sections of this report we have systematically discussed
how to search experimentally for the Higgs boson in the various mass
regimes.  We now take a different tack and ask what we can learn
about the Higgs boson through indirect measurements

\subsection{Unitarity}

In the previous section we discussed  looking for strongly interacting
Higgs bosons through effects which grow with the energy.   However,
models which have cross sections rising with $s$ will eventually
violate perturbative unitarity.  To see this we consider
$2\rightarrow 2$ elastic scattering.  The differential cross section is
\beq
{d \sigma \over d \Omega}={1\over 64 \pi^2 s} \mid {\cal A}\mid^2.
\eeq
Using a partial wave decomposition the amplitude can be written as
\beq
{\cal A}=16 \pi \sum_{l=0}^{\infty} ( 2 l + 1)
P_l(\cos\theta)a_l
\eeq
where $a_l$ is the spin $l$ partial wave and $P_l$ are
the Legendre polynomials.  The cross section
can  now be written as
\beqn
\sigma &=& {8 \pi\over s}\sum_{l=0}^{\infty}
\sum_{l^\prime=0}^{\infty}
(2 l+1) (2 l^\prime + 1) a_l a^*_l\nonumber \\
&&\qquad\cdot
\int_{-1}^1 d \cos\theta P_l(\cos \theta) P_{l^\prime}(\cos \theta)
\nonumber \\
&=& {16 \pi \over s}\sum_{l=0}^{\infty}(2l+1)\mid a_l\mid^2
\eeqn
where we have used the fact that the $P_l$'s are orthogonal.
The optical theorem  gives,
\beq
\sigma={1\over s}Im\biggl[{\cal A}(\theta=0)\biggr]
={16 \pi \over s}\sum_{l=0}^{\infty}(2l+1)\mid a_l\mid^2.
\eeq
This  immediately yields
 the unitarity requirement which is illustrated in
Fig. 37.
\beq
\mid a_l\mid^2=Im(a_l).
\eeq
{}From Fig. 37 we see that one statement of unitarity is the requirement
that
\beq
\mid Re(a_l)\mid < {1\over 2}.
\eeq

As a demonstration of unitarity restrictions we consider the scattering
of longitudinal gauge bosons, $W^+_LW^-_L\rightarrow W^+_LW^-_L$,
which can be found to ${\cal O}(M_W^2/s)$ from the Goldstone boson
scattering of Fig. 34.
We begin by constructing the $J=0$ partial wave in the limit
$M_W^2<<s$ from Eq. \ref{wwwwgold},
\beqn
a_0^0(\omega^+\omega^-\rightarrow
\omega^+\omega^-)& \equiv &{1\over 16 \pi s}
\int^0_{-s}\mid {\cal A}\mid dt
\nonumber \\
&=&- {G_F M_H^2\over 8 \sqrt{2} \pi }
\biggl[2 + {M_H^2\over s-M_H^2}-{M_H^2\over s}\log
\biggl(1+{s\over M_H^2}\biggr)\biggr].
\label{wwscat}
\eeqn

If we go to very high energy, $s >>M_H^2$, then Eq. ~\ref{wwscat}
has the limit
\beq
 a_0^0(\omega^+\omega^-\rightarrow
\omega^+\omega^-)\longrightarrow_{s>>M_H^2}
- {G_FM_H^2\over 4 \pi \sqrt{2}}.
\eeq
Applying the unitarity condition, $\mid Re (a_0^0)\mid< {1\over 2}$ gives
the restriction
\beq
M_H< 860~GeV.
\label{hbound}
\eeq
It is important to understand that this does not mean that the
Higgs boson cannot be heavier that $860~GeV$, it simply means that
for heavier masses perturbation theory is  not valid.
By considering coupled  channels, a slightly tighter bound than
Eq. \ref{hbound} can be obtained.
The Higgs boson therefore plays a fundamental role in the
theory since it cuts off the growth of the partial wave
amplitudes and makes the theory unitary.

We can apply the alternate limit to Eq.~\ref{wwscat} and take the
Higgs boson much heavier than the energy scale.  In this limit\cite{chan}
\beq
a_0^0(\omega^+\omega^-\rightarrow
\omega^+\omega^-)\longrightarrow_{s<<M_H^2}
 {G_F s \over 16 \pi \sqrt{2}}.
\eeq
Again applying the unitarity condition we find,
\beq
\sqrt{s_c}< 1.8~TeV
\label{units}
\eeq
We have used the notation $s_c$ to denote $s$(critical), the scale
at which perturbative unitarity is violated.
Eq. ~\ref{units} is the basis for the oft-repeated statement,``
{\it There must be new physics on the TeV scale}''.
Eq. \ref{units} is telling us that without a Higgs boson, there
must be new physics which restores perturbative unitarity somewhere
below  an energy scale of $1.8~TeV$.

\subsection{Triviality}

Bounds on the Higgs boson  mass have also been  deduced on the grounds
of {\it triviality}.\cite{triv}  The basic argument goes as follows:
Consider a pure scalar theory in which the potential is given
by\footnote{$\mu^2<0$, $\lambda>0$.}
\beq
V(\Phi)=\mu^2\Phi^\dagger \Phi +\lambda (\Phi^\dagger \Phi )^2
\label{potsm}
\eeq
where the quartic coupling is
\beq
\lambda={M_H^2\over 2 v^2}.
\eeq
This is the scalar sector of the Standard Model with no gauge bosons
or fermions.
The quartic coupling runs with renormalization scale $Q$:
\beq
{d \lambda \over dt}={3 \lambda^2\over 4 \pi^2},
\label{lams}
\eeq
where $t\equiv \log(Q^2/Q_0^2)$ and $Q_0$ is some reference
scale. (The reference scale is often
taken to be $v$ in the Standard Model.)
 Eq. \ref{lams} is easily
solvable,
\beqn
{1\over \lambda(Q)}&=&{1\over \lambda(Q_0)}-{3\over 4 \pi^2}
\log\biggl({Q^2\over Q^2_0}\biggr),
\nonumber \\
\lambda(Q)&=& {\lambda(Q_0)\over
\biggl[1-{3\lambda(Q_0)\over 4 \pi^2}\log({Q^2\over Q^2_0})
\biggr]}.
\label{lampol}
\eeqn
{}From Eq. ~\ref{lampol} we see that $\lambda(Q)$ blows up
as $Q\rightarrow \infty$
 (called the Landau pole).  Regardless of how small $\lambda(Q_0)$
is, $\lambda(Q)$ will eventually
become infinite at some large $Q$.
Alternatively, $\lambda(Q_0)\rightarrow 0$ as
$Q\rightarrow 0$ with $\lambda(Q)>0$.  Without the $\lambda\Phi^4$
interaction of Eq.\ref{potsm}
the theory becomes a non-interacting theory at low energy, termed a
trivial theory.  Of course, this picture is valid only if the one
loop evolution equation of Eq.
\ref{lams} is an accurate description of the theory at
large $\lambda$.  For large $\lambda$, however, higher order or
non-perturbative corrections to the evolution equation must be included
to determine if triviality really is a problem for the Standard
Model.

There are many variations on the
triviality theme which attempt to place
bounds on the Higgs mass and we will describe several of them here.
Suppose we consider a theory with only scalars and assume that
there is some new physics at a scale $\Lambda < M_{pl}$.
Then
if we take $\lambda(\Lambda)$ to have its maximum value ($\infty$)
and evolve the coupling down to the weak scale ($v$) we will find
the maximum  allowed value of the Higgs mass,
\beq
\lambda(v)={M_H^2({\rm max})\over 2 v^2}
={4 \pi^2 \over 3 \log({\Lambda^2\over v^2})}.
\label{hlims}
\eeq
Scenarios like this tend to get bounds on the order of $M_H<
{\cal O}(400-1000~GeV)$ as illustrated in Fig. 38.
Since this picture clearly only makes sense for $M_H<\Lambda$,
we have also shown the $\Lambda=M_H^{\rm max}$ curve in
Fig. 38 and we see that for $\Lambda < 1~TeV$, this procedure
for limiting the Higgs boson mass breaks down.  However,
this breakdown leads to the exciting possibility of new
physics in the TeV region  which should be experimentally accessible
at the LHC.
Note that as $\Lambda\rightarrow \infty$,
$M_H^{\rm max}$  quickly reaches its asymptotic value since
the sensitivity of the limit to the cutoff is only logarithmic.
If we apply an arbitrary
cutoff
 of $(\Lambda_{cut}/M_H)> 2 \pi$, we find a limit $M_H < 930~GeV$
from Eq. \ref{hlims}.

Lattice gauge theory calculations have used
similar techniques to obtain
a bound on the Higgs mass.\cite{latbounds}
 One criticism of the previous bounds could
be that it makes no sense to use one loop perturbation theory in
the limit $\lambda\rightarrow \infty$.  Non-perturbative lattice gauge
theory calculations overcome this deficit.  As above, they consider
a purely scalar theory and require that the scalar self coupling
$\lambda$ remain finite for all scales less than $2 \pi M_H$.
This gives a limit\cite{hasen}
\beq
M_H({\rm lattice})< 640~GeV.
\eeq
 The lattice results are relatively
insensitive to the value of the cutoff chosen, as can be seen in Fig.
39.
It is interesting that all of the limits based on the running of the
pure scalar theory tend to be in the $1~TeV$  range, as was the unitarity
bound.

 Of course everything we have done so far is for a theory with only
scalars.  The physics changes dramatically
 when we couple the theory to fermions
and gauge bosons.
Since the Higgs coupling to fermions is proportional to the Higgs
boson mass, the only relevant fermion is the top quark.  When we include
the top quark and the gauge bosons, Eq. ~\ref{lams} becomes\cite{rge}
\beq
\beta_\lambda\equiv
{d \lambda \over d t}={1\over 16 \pi^2}\biggl[
12 \lambda^2 + 6 \lambda h_t^2 - 3 h_t^4-{3\over 2}\lambda(3 g^2
+g^{\prime~2})+{3\over 16} (2 g^4 +(g^2+g^{\prime~2})^2)\biggr]
\label{renorm}
\eeq
where $h_t\equiv \sqrt{2} M_T/v$.  The important physics point is the
opposite signs between the various terms.  For a
heavy Higgs boson , $\lambda>h_t,g,g^{\prime}$, and
\beq
{d \lambda \over d t}\sim{\lambda\over 16 \pi^2}\biggl[
12 \lambda + 6  h_t^2 -{3\over 2}(3 g^2
+g^{\prime~2})\biggr].
\eeq
There is a critical value of $\lambda$ which depends on the top
quark mass,
\beq
\lambda_c\equiv{1\over 8} ( 3 g^2 +g^{\prime~2})-{h_t^2\over 2}
\eeq
 for which  there is no evolution of the scalar
coupling constant.\cite{cab}
 If
$M_H>M_{H}^c\equiv \sqrt{2 \lambda_c} v$ then  the quartic coupling
blows up and the theory is non-perturbative.
If we require that the theory be perturbative (i.e.,
the Higgs quartic coupling be finite) at all energy scales
below some unification scale ($\sim 10^{16}~GeV$)
then an upper bound on the Higgs mass
is obtained as a function of the top quark mass.  For $M_T=170~GeV$
this bound is $M_H < 170~GeV$. \cite{cab}
  If a Higgs boson were found which was
heavier than this bound, it would require that there be some new
physics below the unification scale.
We see that the inclusion of the top quark into the evolution equations
for the scalar coupling has changed the bounds on the Higgs mass
considerably.  Of course this analysis relies on the use of the
one-loop evolution equations.  As yet, there is no lattice bound
on the Higgs mass which incorporates a heavy top quark mass.

\subsection{Vacuum Stability}

A bound on the Higgs mass can also be derived by the requirement that
spontaneous symmetry breaking actually occurs;\cite{linde} that is,
\beq
V(v)< V(0).
\label{vacstab}
\eeq
For small $\lambda$, Eq. ~\ref{renorm} can be solved
to find
\beq
\lambda(Q)=\lambda(Q_0)+\beta_\lambda\log\biggl({Q^2
\over Q_0^2}\biggr),
\eeq
where the small $\lambda$ limit of $\beta_\lambda$ can be found
from Eq. \ref{renorm}.
If we substitute this into the potential, we find
\beq
V(\Phi)\sim\mu^2\Phi^\dagger \Phi +\lambda(Q_0)
(\Phi^\dagger \Phi)^2
+\beta_\lambda
(\Phi^\dagger \Phi)^2
\log\biggl({Q^2
\over Q_0^2}\biggr)  .
\eeq
We can find the minimum of the potential by taking
\beq
{\partial V\over \partial \Phi}\mid_{\Phi=v/\sqrt{2}}=0.
\label{min}
\eeq
Taking the second derivative of the potential and substituting
in the requirement of Eq.~\ref{min} we find the Higgs mass
\beq
M_H^2 = {1\over 2}{\partial^2 V\over \partial\Phi^2}
\mid_{\phi=v/\sqrt{2}}.
\eeq
The requirement of Eq. \ref{vacstab}
 that spontaneous symmetry breaking occurs
gives the famous Coleman-Weinberg bound,\cite{linde}
\beq
M_H^2 > \beta_\lambda v^2 =
{3\over 16 \pi^2 v^2}\biggl(2 M_W^4+M_Z^4-4 M_T^4\biggr).
\eeq
For $M_T> 78 ~GeV$, $\beta_\lambda<0$ and the potential turns over
and is unbounded below.
Of course perturbation theory breaks down for $\Phi\rightarrow
\infty$  and the one loop results are not valid.

 A more careful
analysis\cite{vacbounds} using the $2$ loop renormalization group
improved effective potential\footnote{The renormalization
group improved effective potential sums all potentially large
logarithms, $\log(Q^2/Q_0^2)$.} and the
running of all couplings gives the requirement from vacuum
stability,\cite{sher}\footnote{This limit requires that
the vacuum be stable up to very large scales, $\sim 10^{15}~GeV$.}
\beq
M_H(GeV)> 132 + 2.2(M_T-170)
-4.5\biggl({\alpha_s-.117\over .007}\biggr).
\eeq

We see that when $\lambda$ is small (a light Higgs boson) radiative
corrections become important and lead to a lower limit on the Higgs
boson mass from the requirement of vacuum stability.  If $\lambda$
is large (a heavy Higgs boson) then unitarity and triviality
arguments lead to an upper bound on the Higgs mass.

\subsection{Bounds from Electroweak Radiative Corrections}

The Higgs boson enters into one loop
radiative  corrections in the standard
model and we might hope that precision electroweak measurements
would give some bound on the Higgs mass.
  For example the
$\rho $ parameter gets a contribution from the
Higgs boson\cite{rhoh}\footnote{This result is scheme dependent.  Here
$\rho\equiv M_W^2/M_Z^2\cos^2\theta_W(M_W)$, where $\cos\theta_W$ is
a running parameter calculated at an energy scale of $M_W$.}
\beq
\rho=1-{11 g^2\over 96 \pi^2}\tan^2\theta_W \log
\biggl({M_H\over M_W}\biggr).
\eeq
In fact it is straightforward to demonstrate
that at one loop all  electroweak
parameters have at most a
logarithmic dependance on $M_H$.\cite{abl}  This fact has been
glorified by the name of the ``screening theorem''.\cite{veltman}
  In general,
electroweak radiative corrections involving the Higgs boson take the form,
\beq
g^2\biggl( \log{M_H\over M_W}+g^2 {M_H^2\over M_W^2} ...\biggr).
\eeq
That is, effects quadratic in the Higgs mass are always screened
by an additional power of $g$ relative to the lower order logarithmic
effects and so radiative corrections involving the Higgs
boson can never be large.\cite{einhorn}
We can demonstrate this in terms of the
non-linear model discussed in Section 5.1 where the Higgs boson was removed
from the theory.  One loop corrections to this theory give divergences
which can correctly be interpreted as the $\log(M_H)$ terms of the
Standard Model loops.

{}From precision measurements at LEP of  electroweak observables there is
only the very weak bound on the Higgs boson mass,\cite{lang}
\beq
M_H< 780~ GeV \quad (90\%~ {\rm~ confidence~ level}),
\eeq
although $\chi^2$ fits to the data tend to prefer light Higgs masses.
{}From the LEP data alone, the minimum of the $\chi^2$ fit
to the top quark mass and the Higgs mass is at
 $M_H=60$ GeV,
while if the CDF measurement of the top quark mass is included
the $\chi^2$ minimum occurs at $M_H=120~GeV$.

\section{Problems with the Higgs Mechanism}

In the preceeding sections we have discussed many features of the
Higgs mechanism.  However, many theorists firmly believe that the
Higgs mechanism cannot be the entire story behind electroweak symmetry
breaking.  The primary reasons are:
\begin{itemize}
\item  The Higgs sector of the theory is trivial.

\item  The Higgs mechanism doesn't explain why $v=246~GeV$.

\item  The Higgs mechanism doesn't explain why fermions have
the masses they do.

\item  Loop corrections involving the Higgs boson are quadratically
divergent and counterterms must be adjusted order by order in
perturbation theory to cancel these divergences.  This fine tuning
is  considered by most theorists to be unnatural.

\end{itemize}

In light of the many
objections to the simplest version of the Higgs mechanism
theorists have considered several alternatives
 to the Higgs mechanism for electroweak symmetry breaking.
One proposal, that the electroweak symmetry be broken dynamically
by a mechanism such as technicolor has been discussed at this school by
Appelquist.\cite{aptasi}
Another alternative to the standard model Higgs mechanism is that the
Standard Model becomes supersymmetric.  The electroweak symmetry
is still broken by the Higgs mechanism, but the quadratic
divergences in the scalar sector
are cancelled automatically because of the expanded
spectrum of the theory and so the model is no longer considered to be
unnatural.  In the next section, we  will briefly discuss the phenomenology
of the Higgs bosons occurring in supersymmetric models
and emphasize the similarity of much of the phenomenology to that of the
Standard Model Higgs.
 The theoretical
underpinning of supersymmetric models has been presented at this
school by
Ramond.\cite{ramond}

\section{Higgs Bosons in Supersymmetric Models}

In the standard (non-supersymmetric) model of electroweak
interactions, the fermion masses are generated by
Yukawa terms in the Lagrangian
\beq
{\cal L}=-\lambda_d {\overline Q}_L \Phi d_R-
\lambda_u {\overline Q}_L \Phi^c u_R + h.c.
\eeq
In a supersymmetric theory however, a term proportional to
$\Phi^c$ is not allowed\footnote{$\Phi^c$ cannot be written
as a chiral superfield.} and so another scalar doublet must be
added in order to give the $\tau_3=1$ components of the $SU(2)$
fermion doublets mass.  The mechanism of the symmetry breaking is
very similar to that of the standard model except there are
two Higgs doublets.\cite{ramond,hk}  Before the symmetry
breaking there are two complex scalar $SU(2)$ doublets,
$\Phi_1$ and $\Phi_2$, for
a total of 8 degrees of freedom.
When the spontaneous symmetry breaking occurs,
each scalar obtains a VEV,
 $v_1$ and $v_2$, and the theory is described
in terms of the ratio of VEVs,
\beq
\tan\beta\equiv {v_2\over v_1}.
\eeq
In order that the $W$ mass have the observed value we have the
restriction, $v\equiv \sqrt{v_1^2+v_2^2}=246~GeV$.

 After the spontaneous symmetry
breaking the $W^\pm$ and $Z$ get their longitudinal components
as in the Standard Model
and there are $5$ remaining degrees of freedom.
Supersymmetric models, therefore, have $5$ physical Higgs bosons:  $2$
neutral scalars, $H_1$ and $H_2$, $2$ charged scalars,$H^\pm$ , and
a pseudoscalar, $A^0$. Because of the supersymmetry, the scalar
potential has only one free parameter (unlike the case of the
general two Higgs doublet model where the scalar potential
depends on 6 unknown parameters\cite{hks}).  The masses of the
Higgs scalars can thus be expressed in terms of two parameters
which are conventionally taken to be the mass of the pseudoscalar,
$M_A$, and $\tan\beta$.\footnote{Remember that in the
Standard Model, there was only one free parameter, $M_H$.}
This gives relationships between the masses of the SUSY Higgs
particles.
  The tree level relationships between
the SUSY scalar masses, however, receive large radiative corrections
at one-loop of ${\cal O}(M_T^4/M_W^4)$.\cite{susyrad}
In Fig. 40, we show the mass of the lightest
neutral Higgs boson in terms
of $\tan\beta$ and $M_A$.\cite{bargsusy}
There is an
upper bound to the mass of the lightest Higgs boson  which
depends on the top quark mass through the radiative corrections,
\beq M_{H_1}^2<M_Z^2 +{3 G_F\over \sqrt{2}\pi^2}M_T^4
\log\biggl(1+{{\tilde m}^2\over M_T^2}\biggr)
\eeq
where ${\tilde m}$ is the scale associated with the supersymmetry
breaking, usually taken to be $\sim 1~TeV$.
  For
$M_T\sim 170~GeV$,\cite{susyrad}
\beq
M_{H_1} < 120~GeV.
\eeq
 Hence in a SUSY model, the Higgs mechanism can be
excluded experimentally if a Higgs boson is not found below this
mass scale.
This is in direct contrast to the Standard Model where there is
no prediction for the Higgs boson mass.

There are several relevant features for SUSY phenomenology.  The
first is that the couplings of the neutral scalars to vector
bosons ($V=W^\pm,Z$) are suppressed from those of the standard
model
\beq
g^2_{H_1VV}+g^2_{H_2VV}=g^2_{H VV}(SM)
\eeq
where $g_{HVV}$ is the coupling of the Higgs boson to vector
bosons.
Because of this sum rule, the $WW$ scattering production mechanism
tends not to be as important in SUSY models as in the Standard
Model.

The couplings of the lightest Higgs boson  to the $Z$ are shown in Fig. 41.
The processes $e^+e^-\rightarrow Z H_1$ and
$e^+e^-\rightarrow A^0 H_1$ can be seen from Fig. 41 to be
 complementary in a SUSY model:  they
cannot be simultaneously suppressed.  Using the $2$ modes the $e^+e^-$
machines can cover the SUSY parameter space without holes.  In Fig. 42
we show the range of parameter space which has been excluded by the
ALEPH experiment at LEP.\cite{aleph}  They find:
\beqn
M_{H_1}&>& 43~GeV \nonumber \\
M_{A}&>& 21~GeV.
\label{susylims}
\eeqn
 Of course for a given value of $M_{H_1}$ or $M_A$ there may be a
stronger limit than Eq. \ref{susylims}.

 The second important feature for the phenomenology
of SUSY models is that fermion couplings
are no longer strictly proportional to mass.
The  neutral Higgs boson couplings to fermions are:
\beqn
{\cal L}_{Hff}&=& -{g m_d\over 2 M_W \cos\beta}
{\overline d}  d\biggl(H_2 \cos\alpha - H_1 \sin\alpha\biggr)
\nonumber \\ &&
 -{g m_u\over 2 M_W \sin\beta}
{\overline u}  u\biggl(H_2 \sin\alpha + H_1 \cos\alpha\biggr)
\nonumber \\ &&
 +{i g m_d\tan\beta\over 2 M_W }
{\overline d}  \gamma_5 d A^0
 +{i g m_u\cot\beta\over 2 M_W }
{\overline u}  \gamma_5 u A^0
\label{susycoups}
\eeqn
where $\alpha$ is a mixing angle in the neutral Higgs sector,
\beqn
H_2&=& \sqrt{2}\biggl[(Re\Phi_1^0-v_1)\cos\alpha
+(Re \Phi_2^0-v_2)\sin\alpha\biggr]\nonumber \\
H_1&=& \sqrt{2}\biggl[-(Re\Phi_1^0-v_1)\sin\alpha
+(Re \Phi_2^0-v_2)\cos\alpha\biggr]
\eeqn

At a hadron collider, the neutral SUSY higgs bosons can
be searched for using the same techniques as in the standard
model.  For most choices of the parameter space, gluon fusion
is the dominant production mechanism.  In the Standard
Model, it was only the top quark contribution to gluon
fusion which was important. In a SUSY model, however,
the coupling to the $b$ quark can be important for small values
of $\cos\beta$, as can be seen
from Eq. \ref{susycoups}.

  SUSY models have a rich particle spectrum in
which to search for evidence of the Higgs mechanism.
The various decays such as $H_i\rightarrow \gamma\gamma$, $H^+\rightarrow
l^+\nu$, $A^0\rightarrow \tau^+\tau^-$, etc, are sensitive to
different regions in the $M_A-\tan\beta$ parameter space.
It takes the combination of many decay channels
in order to be able to cover the  parameter space completely with out
any holes.  Discussions of the capabilities of the LHC detectors
to experimentally observe evidence for the Higgs bosons of SUSY
models can be found in the ATLAS\cite{atlas} and CMS\cite{cms}
studies.

\section{Conclusions}

Our current experimental
 knowledge  of the Higgs boson gives only the limits $M_H> 58~GeV$
found from direct searches
and $M_H< 780~GeV$ from precision measurements
at the LEP experiments.  The lower limit
 will be extended to a mass reach on the
order of $80~GeV$ at LEPII.  From here, we must wait until the advent of the
LHC for further limits.  Through the decay $H\rightarrow \gamma \gamma$
and the production process $pp\rightarrow Z l^+l^-$, the LHC will
probe the mass region between $100< M_H< 180~GeV$.  It is an important
question as to whether there will be a hole in
the Higgs mass coverage between the upper reach of LEPII
and the lower reach of the LHC.
Current ideas as to how to look for a Higgs boson in this
mass regime focus on the production mechanism, $pp\rightarrow WH$,
with $H\rightarrow b {\overline b}$.  The efficiency of this
technique, however, depends sensitively on the capabilities of
the LHC to do $b$ tagging at a high luminosity.
For the higher mass region, $180< M_H < 800~GeV$, the LHC will be able
to see the Higgs boson through the gold plated decay mode, $H\rightarrow
ZZ\rightarrow l^+l^-l^+l^-$.

One of the important yardsticks for all current and future accelerators is
their
ability to discover (or to definitively exclude) the Higgs boson
of the Standard Model.
We  hope that at the time of the LHC, we will be able to
probe all mass scales up to $M_H\sim 800~GeV$.  If the Higgs boson is not
  found below this mass scale then we are in the regime where
perturbative unitarity has broken down
and we are led to the
exciting conclusion that there must be new physics beyond the Standard Model
waiting to be discovered.

\end{document}